\documentclass{iopjournal}

\usepackage{graphicx} % Include figure files
\usepackage{dcolumn} % Align table columns on decimal point
\usepackage{xcolor, lipsum}
\usepackage[caption=false]{subfig}
\usepackage{amsmath, amsfonts, amssymb, amsthm, mathrsfs}
\usepackage{mathtools}
\newtheorem{theorem}{Theorem}[section]
\newtheorem{lemma}[theorem]{Lemma}

\newtheorem{definition}{Definition}

\usepackage{bm} % bold math
\usepackage{import} % import sections 
\usepackage{hyperref} % add hypertext capabilities
\usepackage[mathlines]{lineno}% Enable numbering of text and display math 
\usepackage{float}
\usepackage{cite}
\usepackage{wasysym}
\usepackage{ragged2e}
\usepackage{enumitem}
\newcommand{\g}{\gamma}
\renewcommand{\a}{\alpha}
\newcommand{\s}{\sigma}
\renewcommand{\S}{\Sigma}

\renewcommand{\L}{\Lambda}
\renewcommand{\l}{\lambda}

\newcommand{\R}{\mathbb{R}}
\newcommand{\X}{\mathcal{X}}

\renewcommand{\figurename}{Fig.}

\begin{document}

\articletype{Paper} %	 e.g. Paper, Letter, Topical Review...

\justifying

\title{STEM: A Spectral Theory of Ensnarlment Modes}

\author{Chinmayi Subramanya$^{1,2,3*}$\orcid{0009-0004-6403-379X}, Yu Tian$^{1,2,3,4}$\orcid{0000-0002-3872-3971} and Carl D. Modes$^{1,2,3*}$\orcid{0000-0001-9940-0730}}

\affil{$^1$ Max Planck Institute of Molecular Cell Biology and Genetics (MPI CBG), 01307 Dresden, Germany}

\affil{$^2$ Center for Systems Biology Dresden (CSBD), 01307 Dresden, Germany}

\affil{$^3$ Cluster of Excellence, Physics of Life, TU Dresden, 01307 Dresden, Germany}

\affil{$^4$ Max Planck Institute for the Physics of Complex Systems (MPI PKS), 01187 Dresden, Germany}

\affil{$^*$Author to whom any correspondence should be addressed.}

\email{subramanya@mpi-cbg.de, modes@mpi-cbg.de}


\begin{abstract}
    \justifying
    Intertwined spatial networks occur in systems ranging from crystal lattices and polymers to DNA structures and vascular and secretory networks in biological tissues. 
    Topological entanglement may occur within individual components through knotting, or between multiple components through linking.
    We define `ensnarlment' as the topological linkage of distinct spatial networks.
    Two networks are ensnarled when cycles of one network are linked to cycles of the other. The two networks therefore cannot be separated without cutting either of them. 
    The liver provides a striking example: its blood sinusoidal and bile canalicular networks are not individually knotted, yet the two are heavily ensnarled.
    % Their structural interlinking, termed 
    Ensnarlment can carry information about interactions between the networks and can ultimately determine their collective function. 
    However, the features and constraints introduced by such ensnarled architectures cannot be characterized by the connectivity of either network alone. 
    Building on the ensnarlment operator, %introduced by Kramer and Modes,
    %\cite{kramer2023ensnarled} %%% Cannot cite in abstract
    which combines cycle representations and Gauss linking numbers, we develop a spectral framework for quantifying ensnarlment in spatial networks. 
    We investigate the spectral structure through analytical examples of networks of increasing complexity, across ordered and disordered networks.
    We establish the degree of ensnarlment as a global measure of ensnarlment, and ensnarlment centrality as the individual edge contributions within each singular mode. 
    We further establish a rank lower bound on the number of cuts required for unlinking and develop a spectral edge cutting algorithm, distinguishing rank-tight systems from those with positive cut excess. 
    Together, these measures provide complementary descriptions of the extent and distribution of inter-network linkage and the cuts required to eliminate it.
\end{abstract}
%%% Current word count: 256
%%% Word count limit: 300

%simple language about knots vs links especially with regard to polymer entanglement or DNA/chromosome structure. Define ensnarlment at first appearance in the abstract. Give liver as an example up front -- not knotted, heavily ensnarled.

% \newpage
\section{Introduction} \label{sec:introduction}

%%% Outline %%%

% \textcolor{blue}{flow for intro: 
% \begin{enumerate}
%     \item motivate spatial networks
%     \item existing network models aren't spatial enough. give examples
%     \item in many instances, multiple networks can coexist. give examples. in 3d, when multiple networks coexist, they can be interlinked 
%     \item define ensnarlment, describe ensnarlment operator as an adjacency-like operator
%     \item in this paper we look at spectral properties of this operator. motivate spectral graph theory briefly. 
%     \item summarize main results - eigenvalue corresponds to linked cycle interactions; number of nonzero eigenvalues = degree of ensnarlment; ensnarlment centrality 
% \end{enumerate}
% }

Many natural and engineered networks are physically embedded in space, %, and are therefore shaped by local geometries and constraints. 
% Instances include
such as transport and communication systems \cite{crucitti2006centrality, helmy2002small, schintler2006power}, and biological networks such as vasculature \cite{hunt2016asymmetries, papadopoulos2016embedding} and neural circuitry \cite{bullmore2012economy, stiso2018spatial}.
%\yu{[More examples, and individual citations]}
In these systems, the local spatial constraints govern where edges can form and how their lengths, orientations, and arrangements affect network organization and adaptation \cite{barrat2005effects}. 
Geometry therefore provides defining information that is absent from an abstract description of connectivity alone. %\yu{[Make it longer]}
% \cite{barthelemy2011spatial}.
% It is therefore crucial to understand the spatial organization of such networks to understand how they grow and remodel according to local constraints. 

%%% Probably unnecesary 
% Classical network approaches such as Erdős-Rényi and Barabási-Albert models treat connectivity independently of geometric embedding \cite{barabasi1999emergence, erdos1960evolution}. 
% While these models have provided fundamental insights into network organization, they do not explicitly account for spatial constraints, physical proximity, or geometric adaptation. 
% This has motivated the development of spatial network models in which growth, transport, and remodeling are coupled to the underlying geometry of the system. 

Spatial network theory accounts for these effects by coupling graph structure to the positions of nodes and edges \cite{barthelemy2011spatial}. 
Such models may incorporate distance-dependent connection rules, construction costs, transport demands, or spatially localized adaptation. 
Much of the work in this direction has addressed the structure and dynamics of individual embedded networks \cite{hu2013adaptation, murray1926physiological, ronellenfitsch2019phenotypes, katifori2018transport, meigel2018flow, le2024physarum, kramer2023biological}. %\yu{[Add Murray citations (for sure); (maybe) fluctuations; biological backgrounds; more from Karen]}
Multilayer and interdependent network frameworks extend this perspective to systems composed of several interacting graphs, typically representing their coupling through shared entities, inter-layer edges, or functional dependencies \cite{kivela2014multilayer, boccaletti2014structure}. 

A different form of coupling arises when distinct networks coexist within the same spatial domain and are constrained by their joint spatial embedding \cite{kramer2020pare}. 
Such configurations occur, for example, in interpenetrating crystal lattices and polymer networks \cite{batten1998interpenetrating, white2021interpenetrating, bonneau2015high, forgan2011chemical, horner2016knot}, coexisting transport network systems in organs \cite{morales2015versatile, marsh2017architecture, onuki2015new, fowler2018three}, and DNA structures \cite{witz2010dna, michieletto2024kinetoplast, du1995topological}. 
Although these systems have disparate physical origins, they share the possibility that network components are topologically interlinked. 
This linkage is a property of their combined embedding and cannot be inferred from the abstract connectivity of either network alone.
Such topological interlinking is fundamentally different from spatial proximity. 
Two components may be brought arbitrarily close without being linked, whereas linked components cannot be separated by continuous deformations that preserve their connectivity and prevent edge crossings. 
For a pair of closed curves, a standard invariant of this relation is the Gauss linking number \cite{ricca2011gauss}. 
Pairs of networks present a more difficult problem because each network generally contains many overlapping cycles, and any linkage can be distributed across several cycle pairs and their constituent edges. 
Moreover, a network cycle space (equivalently, the first homology group of the network) generically has many possible choices of basis. 
A useful description should therefore combine the topological information carried by cycle linkage with the network structure that distributes these interactions across the network edges.

In this regard, Kramer and Modes (2023) introduced the notion of \textit{ensnarlment}: if edge crossings are not allowed, two spatially embedded network components are “ensnarled” if they cannot be separated in space without removing a subset of edges \cite{kramer2023ensnarled}. 
They introduced an ensnarlment operator 
% constructed from linked cycle interactions. 
% The operator 
that combines topological information from both networks through pairwise Gauss linking numbers between cycles, and further maps onto the networks' edge spaces via cycle-edge incidence matrices. 
%A brief summary of this construction is given in Section~\ref{sec: ensnarlment_operator}.
From the diagonal entries of the corresponding symmetric squared operators, one may assign priority scores to individual edges.
These scores can be further used in a greedy cutting algorithm that iteratively removes high-priority edges until the networks are unlinked \cite{kramer2023ensnarled}. 

However, the spectral structure encoded by the operator remains unexplored, including the number and strengths of its independent interaction modes, the distribution of these modes across the network edges, and how these modes relate to globally minimal unlinking cuts.
% Furthermore, the greedy cutting algorithm does not necessarily find the globally optimum cut set. 
Since the two networks need not contain the same number of edges, the ensnarlment operator is generally non-square, making the singular value decomposition the natural framework for its spectral analysis.
The left and right singular subspaces identify paired edge patterns in the two networks, while the associated singular values quantify the strengths of these interaction modes. 
This decomposition separates the magnitude, multiplicity, and spatial distribution of ensnarlment and remains well defined at the level of singular subspaces when degeneracies occur. 
It therefore provides a route from the global linkage encoded by the operator to mode-resolved descriptions of the coupled architecture.

In our study, we use this singular structure to develop a quantitative spectral framework for characterizing ensnarlment. 
%%% Probably unnecessary 
% Traditionally, spectral graph theory provides a powerful framework for understanding the structural properties of complex networks. 
% By analyzing the eigenvalues and eigenvectors of a graph's adjacency matrix, spectral methods reveal critical information about the graph's connectivity, clustering, and overall topology. 
% The spectrum of a graph encodes essential features such as the network's robustness, the presence of communities, and the importance of individual nodes within the structure. 
% These spectral properties are key to identifying hidden patterns within the graph and can help predict the behavior of the network under various conditions. 
% \cite{chung1997spectral}
% We borrow this idea to uncover ensnarled network interaction patterns reflected by the eigenvalue and eigenvector spectra of the ensnarlment operator.
%
Through a range of ensnarled network archetypes with varying complexity, we show that the spectrum of the ensnarlment operator admits a direct structural interpretation. 
The nonzero singular values quantify the strengths of independent ensnarlment modes arising from linked cycle interactions.
The number of these independent modes defines the degree of ensnarlment.
We further define ensnarlment centrality for each singular subspace, which quantifies the contribution of each edge to the corresponding ensnarlment mode.
We then prove that the degree of ensnarlment provides a lower bound on the number of edge removals required for unlinking, and we develop a singular-subspace-guided edge cutting algorithm. 
When the algorithm attains this bound, the resulting cut set is certified to be globally minimal, and the system is classified as rank-tight. 
Systems whose minimum cut set size exceeds the bound are instead characterized by a positive cut excess.
Our results establish a unified spectral description of ensnarlment that connects the global organization of linked modes to their distribution across individual edges and to the cuts required for separation. 
% As the framework can be applied to both model systems and experimentally reconstructed spatial networks, it provides a common basis for comparing ensnarlment across interpenetrating materials and biological network architectures. 
% It further opens directions for investigating how inter-network linkage influences transport, mechanical response, growth, and remodeling in multicomponent systems.

\section{Theoretical overview}\label{sec:theoretical_overview}
% \cs{add some fluff here}
This section develops the mathematical framework for quantifying the ensnarlment of two spatial networks. 
We first summarize the construction of the edge space ensnarlment operator from the networks' cycle structures and pairwise linking, and then use its spectral properties to characterize the number, strength and spatial organization of the resulting ensnarlment modes. 

\subsection{The ensnarlment operator} \label{sec: ensnarlment_operator}

Here we utilize graph theoretical metrics to provide a summary of the ensnarlment matrix framework developed by Kramer and Modes \cite{kramer2023ensnarled}.
Readers who are already familiar with these concepts and methodologies may choose to skip this section.
In the following paragraphs, we outline the foundational principles and relevant metrics that set up the ensnarlment operator.

Each network in the ensnarled system is set up as a graph $\mathscr{G}_i =(\mathcal{V}_i,\mathcal{E}_i)$ with $\mathcal{V}_i$ vertices and $\mathcal{E}_i$ edges. 
Cycles of the graph are defined as the sequence of edges that ends with the same vertex as it begins, which transverses any vertex exactly once. 
Each cycle of the graph may be represented as a combination of basis cycles. 
Such a cycle basis set can be constructed by identifying fundamental cycles of the graph by adding any edge to its spanning tree: a sub-graph of $\mathscr{G}$ which contains all vertices $\mathcal{V}$ and $\mathcal{V}-1$ edges of $\mathscr{G}$. 
The number of fundamental cycles for a spanning tree is the number of edges of $\mathscr{G}$ that it does not contain: 
\begin{equation}
\zeta = \mathcal{E} - \mathcal{V}+1.
\end{equation}
$\zeta$ is referred to as the \textit{cyclomatic number} or the \textit{nullity} of $\mathscr{G}$, which represents the dimension of the graph's cycle space. 

Each basis cycle can be represented as a closed spatial curve with parametrization $\g_i$. 
The linking state of two such spatial curves can be computed using the Gauss linking integral:
\begin{equation} \label{eqn: gli}
    l(\g_1,\g_2) = \frac{1}{4\pi} \int_{\g_1} \int_{\g_2} (d \bm{r}_1 \times d \bm{r}_2) \frac{ \bm{r}_1 -  \bm{r}_2}{|\bm{r}_1 -  \bm{r}_2|^3}.
\end{equation}

The linking state of the ensnarled system is represented by the
% The pairwise linking states of the cycle bases of $\mathscr{G}_1$ and $\mathscr{G}_2$ are collected in the 
linkage matrix $L$, whose entries are
\begin{equation}
     L_{ij} = l(c_{1i},c_{2j}),
\end{equation}
where $l(c_{1i},c_{2j})$ is the linking number between basis cycles $c_{1i}$ and $c_{2j}$ obtained from Eq.(\ref{eqn: gli}).

% \textcolor{red}{
% The linking state of the ensnarled system is represented by the pairwise linking numbers between cycles of $\mathscr{G}_1$ and $\mathscr{G}_2$.
% Let 
% \begin{equation}
%     c_1 = \sum_{i=1}^{\zeta_1} \eta_{1i} c_{1i} \ , \qquad 
%     c_2 = \sum_{j=1}^{\zeta_2} \eta_{2j} c_{2j}
% \end{equation}
% where $ \bm{\eta}_1 \in \mathbb{Z}^{\zeta_1}$ and $ \bm{\eta}_2\in \mathbb{Z}^{\zeta_2}$ are coefficient vectors specifying the multiplicity and orientation of the basis cycles. 
% The total linking number of these cycle combinations is 
% \begin{equation}
%     l(c_1,c_2) = \sum_{i,j}^{\zeta_1,\zeta_2} \eta_{1i} \eta_{2j} \  l(c_{1i},c_{2j})
% \end{equation}
% Or in terms of the linking number operator:
% \begin{equation}
%     l(c_1,c_2) = \bm{\eta}_1^T \cdot L \cdot \bm{\eta}_2
% \end{equation}
% }

% The total linking state of the ensnarled system is represented by the sum of all linking states of cycles $c_1$ of $\mathscr{G}_1$ with cycles $c_2$ of $\mathscr{G}_2$:
% \begin{equation}
%     l(c_1,c_2) = \sum_{i,j}^{\zeta_1,\zeta_2} \s_{1i} \s_{2j} \  l(c_{1i},c_{2j})
% \end{equation}
% Or in terms of the linking number operator:
% \begin{equation}
%     l(c_1,c_2) =  \sigma}_1^T \cdot L \cdot  \s}_2
% \end{equation}

Further, mesh matrices $M_i$ are used to indicate which edges are contained in which cycles, defined as 
\[
     % M_{e_i c_i}
     \left(M_i\right)_{e c} = \begin{cases}
        \pm 1, & \text{if cycle $c$ contains edge $e$ in network $i$}, \\
        0, & \text{otherwise}.
    \end{cases}
\]
The sign of $\left(M_i\right)_{e c}$ indicates whether cycle $i$ traverses edge $e$ along or against its assigned orientation.
% \yu{[...We need text to explain the value of $1$ or $-1$...]}

These are factorized from $L$ to obtain the \textit{ensnarlment operator} $\L$:
\begin{equation}
     L =  M_1^T \cdot \L \cdot  M_2.
\end{equation}

The ensnarlment operator $\L$ is constructed from the generalized inverses of $M_i$ 
\cite{kramer2023ensnarled}:
\begin{equation}
    \L = {M_1^+}^T \cdot  L \cdot  M_2^+.
\end{equation}

The mesh matrices have full column rank, equal to the cyclomatic number of the respective graph. 
The generalized inverse thus takes the form of the left inverse:
\begin{equation}
     M_i^+ = (M_i^T  M_i)^{-1}  M_i^T,
\end{equation}
where $M_i^T  M_i =  G_i$ is the Gram matrix of $M_i$.

% The symmetric inner products of the ensnarlment operator $\L$ are defined as the \textit{symmetric ensnarlment operators} which operate on the edge spaces of the graphs respectively. 
% \begin{equation}
%     S_1 = \L \L^T, \ S_2 = \L^T \L
% \end{equation}

% These operators indicate which edges are critical to the linkage, and can further be used to determine a cut set of edges to achieve unlinking of the two networks. \cite{kramer2023ensnarled} 

% Spectral decomposition of these symmetric ensnarlment operators result in a set of eigenvalues and eigenvectors in the edge spaces of the respective graphs. We will henceforth denote the spectrum $\s$ as the SVD of $\L$, with the left and right singular vectors denoting the eigenvectors on each edge space. 

\subsection{Spectral characterization} \label{sec:spectral_characterization}

Although $\L$ is constructed from cycle linking information rather than direct graph connectivity, it operates between the edge spaces of the two networks. 
It may therefore be interpreted as a weighted adjacency-like operator whose entries encode edgewise interactions induced by linked cycles rather than physical edge incidence. 
This interpretation motivates the application of spectral methods to the ensnarled system. 
{We also provide an alternative view of the ensnarlment operator from the (signed) graph Laplacian in Sec.~\ref{sec:laplacian}.}

Because $\L$ maps between two generally distinct edge spaces, it need not be square. 
Its natural spectral representation is therefore the singular value decomposition (SVD),
\begin{equation}
    \L=U\Sigma V^T,
\end{equation}
where the singular values $\s_i$ quantify the strengths of the independent coupled interactions. 
The left and right singular vectors define the corresponding orthogonal modes on the edge spaces of the respective networks, and their components quantify the participation of individual edges in each mode.
For a degenerate singular value, however, the individual singular vectors are not unique under orthogonal rotations within the degenerate subspace. 
We therefore define edge participation at the level of the complete singular subspace.

\begin{definition}[Ensnarlment centrality] \label{def: ensnarlment_centrality}
    Let $\bm{u}_i$ and $\bm{v}_i$ denote the left and right singular vectors associated with $\s_i$. 
    For a singular value $\s$, the left and right ensnarlment centralities of edges
    $e \in \mathcal{E}_1$ and $f \in \mathcal{E}_2$ are defined by
    \[
        c^{(1)}_e(\s) = \sum_{i:\s_i=\s} \left| (\bm{u}_i)_e \right|^2,
        \qquad
        c^{(2)}_f(\s) = \sum_{i:\s_i=\s} \left| (\bm{v}_i)_f \right|^2 .
    \]
\end{definition}

% \begin{definition}[Principal ensnarlment centrality]
%     Let $\bm{u}_i$ and $\bm{v}_i$ denote the left and right singular vectors associated with $\s_i$. 
%     For the principal singular value $\s_1$, the left and right principal ensnarlment centralities of edges
%     $e \in \mathcal{E}_1$ and $f \in \mathcal{E}_2$ are defined by
%     \[
%         c^{(1)}_e(\s_1) = \sum_{i:\s_i=\s_1} \left| (\bm{u}_i)_e \right|^2,
%         \qquad
%         c^{(2)}_f(\s_1) = \sum_{i:\s_i=\s_1} \left| (\bm{v}_i)_f \right|^2
%     \]
%     % These quantities measure the participation of each edge in the left and right principal singular subspaces associated with $\s_1$. 
%     % For a nondegenerate principal singular value, they reduce to the squared components of the corresponding left and right singular vectors.
% \end{definition}
These quantities measure the participation of each edge in the corresponding 
% principal 
singular subspace. 
They are the diagonal entries of the orthogonal projectors onto these subspaces and are therefore invariant under rotations of the singular vector bases.
If $\s$ is nondegenerate, they reduce to the squared components of the corresponding left and right singular vectors.

Principal ensnarlment centrality, corresponding to $\s = \s_1$, is analogous in a broad spectral sense to graph eigenvector centrality, in which %components of 
the principal eigenvector of the adjacency matrix assigns centrality scores to nodes \cite{bonacich1972factoring}.
Here, however, the scores are defined on the edges of the two ensnarled networks and are resolved by the ensnarlment mode. 
% Particularly, ensnarlment centrality does not assume that edges with large scores occupy the geometric center of the system. 

The number of independent nonzero modes is given by
\begin{equation}
    r_{\mathrm{ens}} = \operatorname{rank}(\L) =  \left| \{ i : \s_i > 0 \} \right|,
\end{equation}
which defines the ensnarlment rank. 
We develop this rank-based characterization and its relation to the linkage matrix formally in Sec.~\ref{sec:degree_ensnarlment}.

% For a nondegenerate singular value, the participation of an edge is given by the squared magnitude of the corresponding singular vector component. 
% For a degenerate singular value, however, the individual singular vectors are not unique under orthogonal rotations within the degenerate subspace. 
% Edge participation must therefore be assigned to the complete singular subspace rather than to an individual vector.

% \yu{[...We need to make it clear that it is a special case of our analysis...]}
The SVD applies to arbitrary pairs of networks. 
In  order to obtain a more tractable analytical formulation, we begin by considering a symmetric special case of the general framework, in which the two networks are identical and represented in corresponding edge and cycle coordinates. 
Setting $M_1=M_2=M$ yields such a simpler formulation in cycle space. 
Let $G=M^TM$, with $M$ of full column rank. 
% admit a simpler formulation in cycle space. 
% Consider two identical networks represented in corresponding edge and cycle coordinates, such that $M_1=M_2=M$, and let $G=M^TM$, with $M$ of full column rank. 
When the linkage matrix $L$ is symmetric, $\Lambda=MG^{-1}LG^{-1}M^T$ is also symmetric, and its singular values are the absolute values of its eigenvalues. 
Applying Sylvester's determinant identity reduces its nonzero eigenspectrum from edge space to cycle space:
\begin{equation}
    \operatorname{spec}_{\ne 0}(\L) = \operatorname{spec}_{\ne 0}(LG^{-1}).
\end{equation}
The derivation is given in Appendix~\ref{app:a1}. 
For the further special case $L^2=I$, $LG^{-1}$ is similar to $(LG)^{-1}$, and hence
\begin{equation} \label{eqn: sylvester_reduction}
    \operatorname{spec}_{\neq 0}(\L) = \frac{1}{\operatorname{spec}_{\neq 0}(LG)}.
\end{equation}

These relations provide the basis for the analytical spectra derived below. 
The following sections examine how the resulting singular values and singular subspaces encode the topology and spatial organization of linked cycle interactions across different ensnarled network classes.

\section{Spectral structure of ensnarled networks} \label{sec:spectral_properties}

% To explore the spectral properties of the symmetric ensnarlment operators $S_i$, we analyze a range of networks with varying complexity, from simple planar graphs to more intricate three-dimensional systems. 
% The goal of this section is to understand how the eigenvalue spectra and corresponding eigenvectors reveal the structural complexity inherent in these networks. 
% We begin with simple networks, where we visualize how eigenvectors correspond to ensnarled edges, providing insight into the hierarchical structure of the system. 
% Next, we examine larger three-dimensional lattice networks which exhibit high degrees of symmetry and more complex connectivity. 
% Finally, we investigate random Voronoi networks that introduce a more stochastic and irregular structure. 
% We present the eigenvalue spectra and eigenvector patterns for each class of networks.

In this section, we investigate the spectral properties of the ensnarlment operator across a range of network archetypes. We address two complementary questions. First, what structural information is encoded in the nonzero singular values of the operator? Second, how do the corresponding singular vectors characterize the organization of the ensnarled system? To answer these questions, we analyze a sequence of increasingly complex network classes, beginning with minimally linked systems and progressing through periodic lattice networks to irregular random Voronoi networks. This progression allows us to distinguish general spectral features from those arising from particular geometries.

\subsection{Minimal Linked Systems}
\begin{figure}[h]
    \centering
    \includegraphics[width=0.7\linewidth]{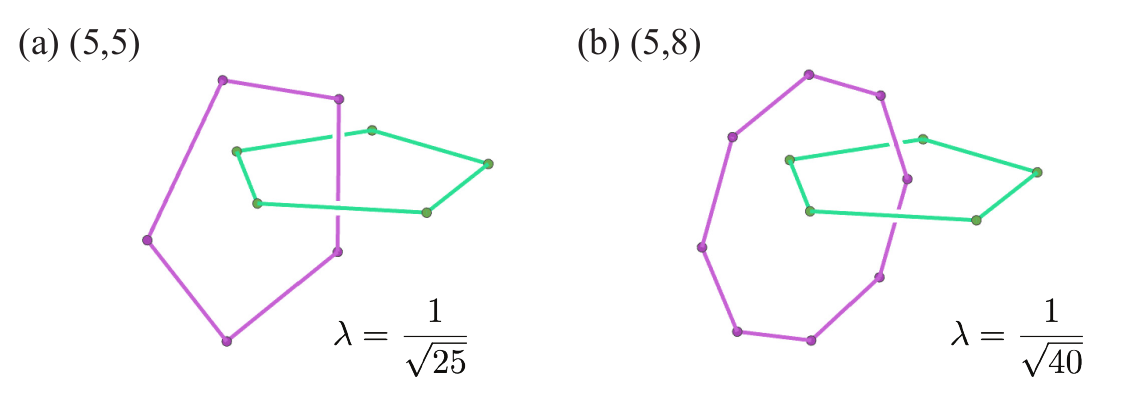}
    \caption{Hopf-linked cycles and their corresponding singular values}
    \label{fig:hopf}
\end{figure}

We first look at Hopf-linked loops. The linkage matrix $L$ for such a system takes the form of the identity matrix. 
This allows us to explore the individual contributions from the mesh matrices. 
% We first consider a single Hopf-link and further generalize into multiple disconnected cycle components of individual graphs, Hopf-linked with each other 
(\figurename~\ref{fig:hopf}). 

Consider two Hopf-linked cycle graphs of path lengths $|\mathcal{E}_1| = n$ and $|\mathcal{E}_2| = m$. The linkage matrix $L$ here is the $1 \times 1$ identity matrix $I_1$; 
the mesh matrix of network $i$ is the all-ones column vector $M_i = \bm{j}_{|\mathcal{E}_i|}$.
The generalized inverse constructions are the row vectors $M_i^+ = 1/|\mathcal{E}_i| \ \bm{j}_{|\mathcal{E}_i|}^T$.

The final construction of the ensnarlment operator $\L \in \mathbb{R}^{n \times m}$ is
\[
    \L = \frac{1}{nm} \ J_{n,m} \ , 
\]
where $J_{n,m}$ is the $n \times m$ all-ones matrix. 
To construct the SVD of $\L$ we use the symmetric ensnarlment operator $\L \L^T \in \mathbb{R}^{n \times n}$,
\[
    \L \L^T = \frac{m}{(nm)^2} \ J_{n,n} \ .
\]

The characteristic polynomial $p_{\L \L^T} (\l)$ is then written to compute the singular values. 
\[
    p_{\L \L^T} (\l ) = \l^{n-1} \left( \l - \frac{1}{nm} \right) \frac{m^n}{(nm)^{2n}}.
\]
This gives $n-1$ zero singular values and one nontrivial singular value (\figurename~\ref{fig:hopf}):
\[
    \sigma = \sqrt{\lambda} = \frac{1}{\sqrt{nm}}.
\]

% This result can be generalized to the case where graphs $G_1$ and $G_2$ consist of disjoint cycles that are Hopf-linked: each link produces a singular value $\sigma = \frac{1}{\sqrt{nm}}$, where $n,m$ are the number of edges in the corresponding parent cycles respectively 
% (\figurename~\ref{fig:hopf}).

This provides a direct interpretation of the singular values in terms of the lengths of the cycles participating in an individual isolated Hopf link. 
The corresponding singular vectors assign equal magnitude components to all edges of each cycle, reflecting the equivalent participation of those edges in the link. 
This minimal example therefore establishes the spectral structure of isolated linked cycle interactions. 
We next ask how this picture changes when neighboring cycles within each network share edges. 

\subsection{Ensnarled ladders} \label{sec: ladders}

We increase the network complexity by extending the linked cycles along one spatial direction to form ladder graphs. 
We define the ladder period $p$ as the number of translational repeats along this direction. 
A ladder of period $p$ contains $2(p+1)$ vertices, $3p+1$ edges, and $p$ independent 4-cycles.
These cycles are independent in the cycle space sense, but neighboring cycles are not edge-disjoint, since they share a rung.
% Each basis cycle in one ladder remains Hopf-linked to the corresponding basis cycle in the other. 
Increasing the ladder period by one therefore introduces one additional linked cycle interaction while simultaneously coupling it to the existing modes through these shared edges. 

\begin{figure}[t]
    \centering
    \includegraphics[width=\linewidth]{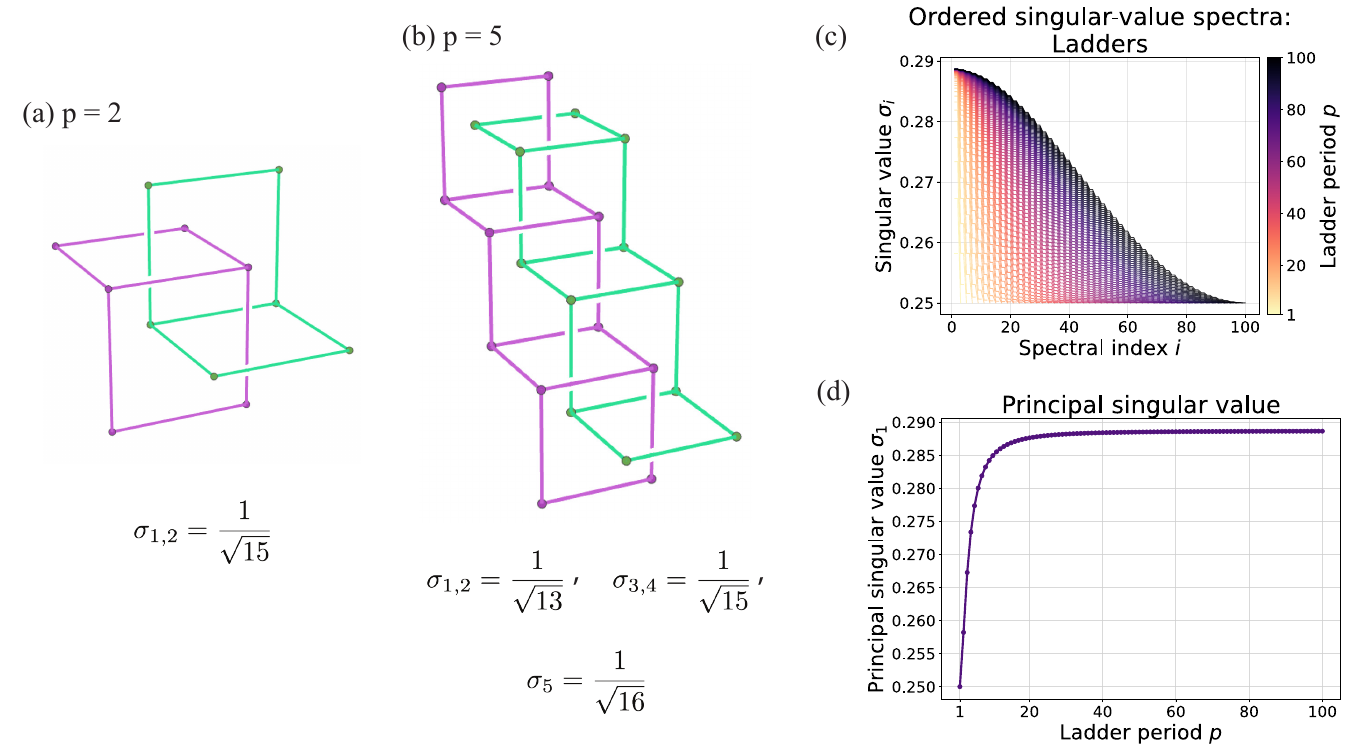}
    \caption{Spectral properties of ensnarled ladders. (a-b) Ladders of period $p=2$ and $p=5$ and their respective singular values in descending order $\s_1 \geq \s_2 \geq \ldots \geq \s_p$. (c-d) Ordered spectra and convergence of the principal singular value over 100 periods.}
    \label{fig:ladders_singvals}
\end{figure}

We consider identical ensnarled ladders in which each 4-cycle of one graph is Hopf-linked to the corresponding 4-cycle in the other graph
(\figurename~\ref{fig:ladders_singvals}a,b).
With a consistent choice of cycle orientations, the linkage matrix $L$ is diagonal with alternating entries, while the Gram mesh matrices $G$ are tridiagonal Toeplitz matrices:
\[
    L = \operatorname{diag}(1, -1, 1, -1, \ldots),
    \qquad
    G = \operatorname{tridiag}(4, -1, -1).
\]
The singular value spectrum of $\L$ can then be obtained from the eigenspectrum of $LG$ through Eq.(\ref{eqn: sylvester_reduction}). 
To evaluate this spectrum, we exploit the symmetry and tridiagonal nature of the system. 
Defining $K=\operatorname{tridiag}(0,-1,-1)$, such that $G=4I+K$, $(LG)^2$ takes the form $16I - K^2$. 
The spectrum can now be computed using the spectrum of $K^2$.

% \[
%     K = tridiag(0,-1,-1),
%     \qquad
%     G = 4I + K.
% \]
% The alternating structure of $L$ gives
% \[
%     L^2 = I,
%     \qquad
%     LKL = -K,
% \]
% and therefore
% \[
%     (LG)^2 = 16I-K^2.
% \]

The eigenvalues of a tridiagonal Toeplitz matrix have the closed form \cite{smith1985numerical}
\begin{equation} \label{eqn: toeplitz_tridiagonal_spectrum}
    \operatorname{spec}_{\neq 0}(\operatorname{tridiag}(a,b,c))
    =
    a + 2 \sqrt{bc} \cos \left( \frac{k\pi}{n+1} \right),
    \qquad
    k = 1,2, \ldots, n.
\end{equation}
Consequently,
\[
    \operatorname{spec}_{\neq 0} (LG)^2 = 16-4 \cos^2 \left( \frac{k\pi}{p+1} \right), \qquad k = 1,2, \ldots, p.
\]
From Eq.(\ref{eqn: sylvester_reduction}), the singular values of $\L$ are the inverse square roots of these eigenvalues. A full derivation is given in Appendix~\ref{app:b1}.

Writing each distinct singular value once, the resulting singular spectrum is
\begin{gather}
    \s_{\neq 0} (\L) = \frac{1}{\sqrt{14 + 2 \cos \theta}}, \nonumber
    \\
    \text{where}
    \theta = \begin{cases}
        \displaystyle{\frac{(2j-1)\pi}{p+1}}, & \text{even } p,
        \\[5pt]
        \displaystyle{\frac{2(j-1)\pi}{p+1}}, & \text{odd } p,
    \end{cases} \qquad
    j = 1,2, \ldots, \lceil p/2 \rceil .
    \label{eqn: ladder_spectra}
\end{gather}

The spectrum has a parity dependent degeneracy structure. 
For even $p$, every singular value $\s$  is two-fold degenerate. 
For odd $p$, all singular values are two-fold degenerate except for a single nondegenerate singular value $\s = 1/4$.

This structure follows from the symmetry
\[
    \cos^2 \left( \frac{k\pi}{p+1} \right) = \cos^2 \left( \frac{(p+1-k)\pi}{p+1} \right).
\]
The complementary indices $k$ and $p+1-k$ therefore produce the same singular value. 
For even $p$, all indices form distinct complementary pairs. 
For odd $p$, however, there is one fixed point,
\[
    k=\frac{p+1}{2}.
\]
At this index,
\[
    \cos \left( \frac{k\pi}{p+1} \right) = 0,
    \qquad
    \lambda_k \left( (LG)^2 \right) = 16,
\]
and hence
\[
    \s = \frac{1}{\sqrt{16}} = \frac{1}{4}.
\]
% Since $16$ is the largest eigenvalue of $(LG)^2$, this nondegenerate mode corresponds to the smallest singular value of $\L$.

The closed form spectrum in Eq.(\ref{eqn: ladder_spectra}) further shows that the extremal singular values converge to size-independent bounds with increasing ladder period. 
For both even and odd $p$,
\begin{equation} \label{eqn: ladder_singvals_limit}
    \s_{\mathrm{principal}} \rightarrow \frac{1}{\sqrt{12}},
    \qquad
    \s_{\mathrm{minor}} \rightarrow \frac{1}{\sqrt{16}} = \frac{1}{4},
    \qquad
    \text{as } p \rightarrow \infty.
\end{equation}
Thus, increasing the size of the ladder does not lead to an indefinite increase in the strength or range of its ensnarlment modes. 
Instead, the spectrum saturates to fixed upper and lower bounds 
(\figurename~\ref{fig:ladders_singvals}d).

This saturation may be interpreted in terms of the effectively one dimensional growth of the ladders. 
As the ladders are extended, the strongest ensnarlment modes increasingly probe edges that are interior along the elongation axis. 
However, these edges do not become progressively enclosed in the two transverse directions, since the network size is unchanged along these directions. 
The degree to which an edge can become `interior' to the ensnarled structure is therefore limited even as the ladder lengths increase.

The corresponding singular vectors can also be obtained analytically. 
The eigenvectors of the tridiagonal Toeplitz matrix $K$ are the discrete sine modes \cite{smith1985numerical}
\begin{equation}
    (\phi_k)_j = \sqrt{\frac{2}{p+1}}
    \sin \left( \frac{jk\pi}{p+1} \right),
    \qquad j, k=1,2, \ldots, p.
\end{equation}

Since $G = 4I + K$, these modes also diagonalize $G$ with eigenvalues 
\[
    g_k = 4 - 2 \cos \left( \frac{k \pi}{p+1} \right).
\]

The alternating linkage matrix maps each mode onto its complementary mode, 
\[
    L \bm{\phi}_k = \bm{\phi}_{p+1-k}.
\]

Each two-fold degenerate singular value is associated with the two-dimensional subspace spanned by the complementary sine modes $\bm{\phi}_k$ and $\bm{\phi}_{p+1-k}$. 
Thus, although the individual singular vectors within a degenerate pair are not uniquely defined, the corresponding singular subspace has a well-defined sine modes. % sinusoidal structure.

\begin{figure}[h]
    \centering
    \includegraphics[width=0.75\linewidth]{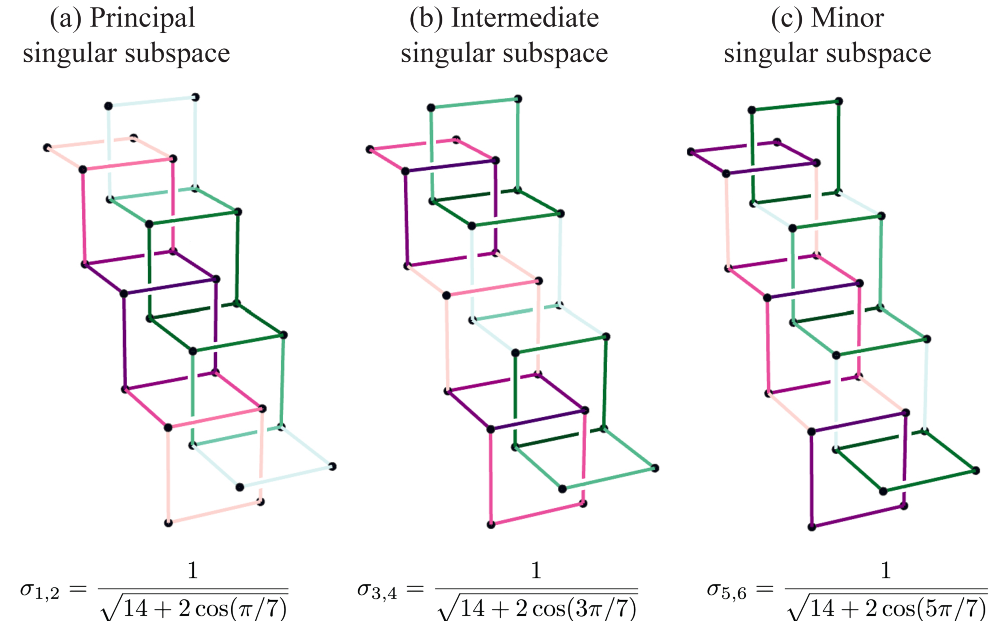}
    \caption{Spatial hierarchy stratified by the singular vectors for ladders of period $p=6$. (a) Internal edges dominate the principal singular subspace and (b-c) peripheral edges dominate lower singular subspaces.
    The two color families distinguish the two networks. 
    Edge darkness encodes the magnitude of the corresponding component of the unit-normalized singular vector: darker edges have larger component magnitudes and participate more strongly in that singular mode. 
    Color intensity should therefore be interpreted relative to other edges within the same displayed vector.}
    \label{fig:ladder_singvecs}
\end{figure}

We visualize each singular subspace using the edgewise ensnarlment centralities defined in Definition~\ref{def: ensnarlment_centrality}.
Edges with high principal ensnarlment centrality are concentrated in the interior of the system, while subspaces associated with smaller singular values 
% increasingly 
emphasize peripheral edges  
(\figurename~\ref{fig:ladder_singvecs}).

\subsection{Ensnarled grids} \label{sec:grids}

\begin{figure}[h]
    \centering
    \includegraphics[width=\linewidth]{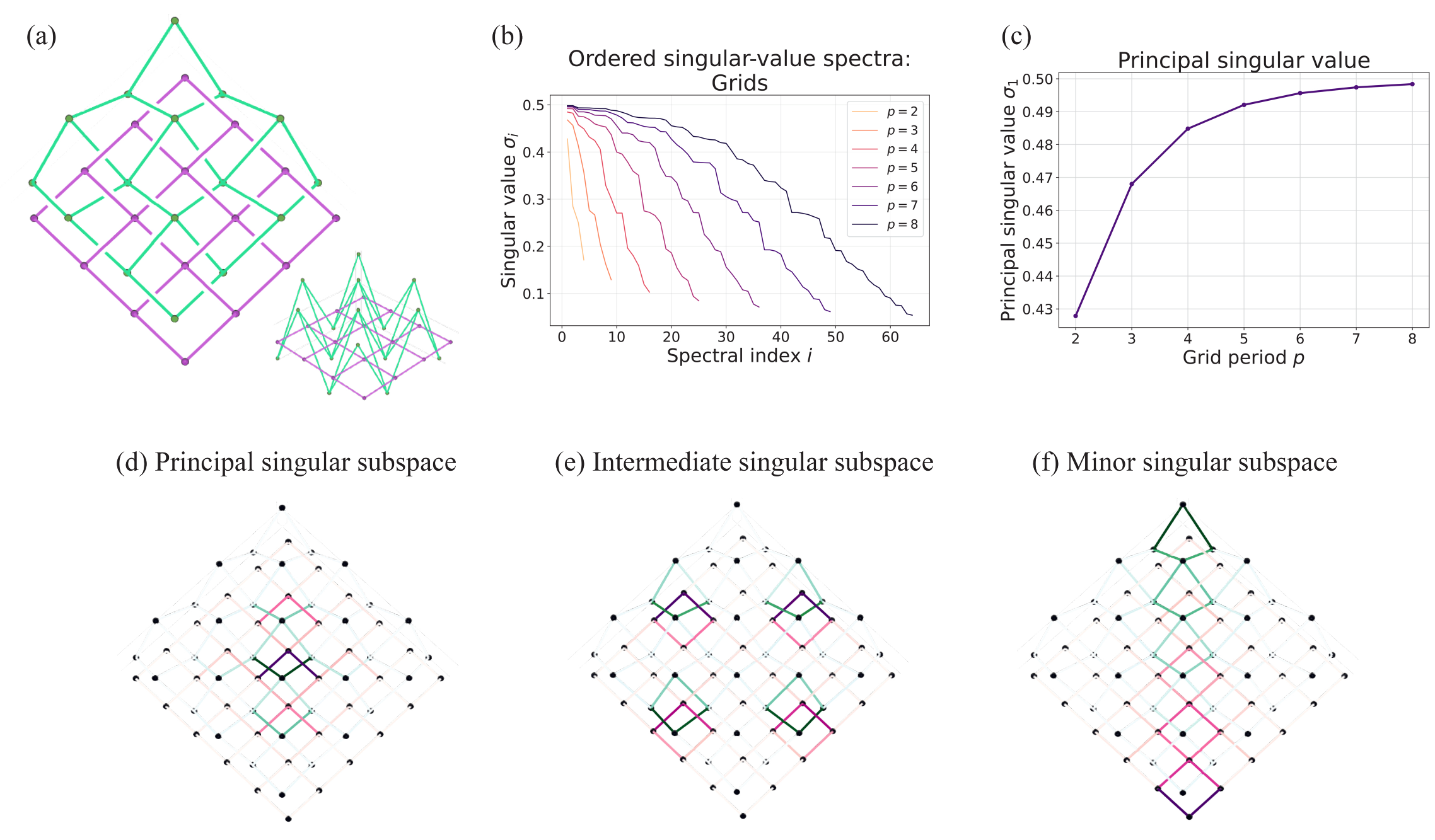}
    \caption{Spectral properties of ensnarled grids. (a) A pair of ensnarled $3 \times 3$ grids and their spatial embeddings. (b-c) Ordered spectra and convergence of the principal singular value over periods $p=2 \ldots 8$. (d-f) Spatial hierarchy stratified by the singular vectors for a pair of ensnarled $5 \times 5$ grids. (d) Internal edges dominate the principal singular subspace and (e-f) peripheral edges dominate lower singular subspaces.
    The two color families distinguish the two networks. 
    Edge darkness encodes the magnitude of the corresponding component of the unit-normalized singular vector: darker edges have larger component magnitudes and participate more strongly in that singular mode. 
    Color intensity should therefore be interpreted relative to other edges within the same displayed vector.}
    \label{fig:grids}
\end{figure}

To investigate how the size independent saturation behavior shown in Eq.(\ref{eqn: ladder_singvals_limit}) changes when the network grows along an additional dimension, we next consider ensnarled grid networks. 
The grids are topologically planar, but one grid is given a corrugated 3D embedding while the other remains flat 
(\figurename~\ref{fig:grids}a).
This produces a controlled ensnarled geometry in which each cycle participates in an ensnarlment mode. 
We consider pairs of equivalent square grids with linear size $p$ such that increasing $p$ extends the networks along both directions of the grid. 
Because the $p \times p$ grid contains $p^2$ independently linked cycle pairs, the linkage matrix is full rank and the ensnarlment rank follows exactly as $r_\mathrm{ens} = p^2$.

The singular value spectra again approach a size independent limit, with the principal singular value converging as 
\[
    \s_\mathrm{principal} \rightarrow \frac{1}{2} 
    \qquad 
    \text{as } p \rightarrow \infty.
\]

This limiting value follows from the regular interior structure of the grids, as shown 
in Appendix~\ref{app:b2}. 
Away from the boundaries, the Gram and linkage operators are translationally invariant up to the alternating linkage signs and therefore admit a two dimensional plane wave representation. 
The resulting singular value function is bounded above by $1/2$, with the bound attained by modes having equal wave vector components along the two grid directions. 
As $p$ increases, the finite grid resolves these interior modes increasingly closely, while boundary contributions become relatively less important, causing $\s_\mathrm{principal}^{(p)}$ to approach $1/2$ 
(\figurename~\ref{fig:grids}c).

This grid limit is higher than the corresponding ladder limit of $1/\sqrt{12}$. 
This difference does not follow from dimensionality alone; it arises from the specific coupling structure encoded jointly by the Gram and linkage operators, which supports collective modes extending along both grid directions.
% but from the two dimensional Gram and linkage structure, which permits collective bulk modes extending along both grid directions. 
Geometrically, an increasing fraction of the edges lies in the interior and is surrounded by linked cycles along two directions. 
The range of these local interactions nevertheless remains bounded, so the principal singular value continues to saturate rather than growing indefinitely. 
The singular vectors further resolve this spatial organization via ensnarlment centrality: the dominant modes are concentrated within the interior of the ensnarled structure, whereas modes lower in the spectrum place comparatively greater weight near its boundaries 
(\figurename~\ref{fig:grids}d-f).

\subsection{Three dimensional ensnarled networks}

\begin{figure}[h]
    \centering
    \includegraphics[width=0.8\linewidth]{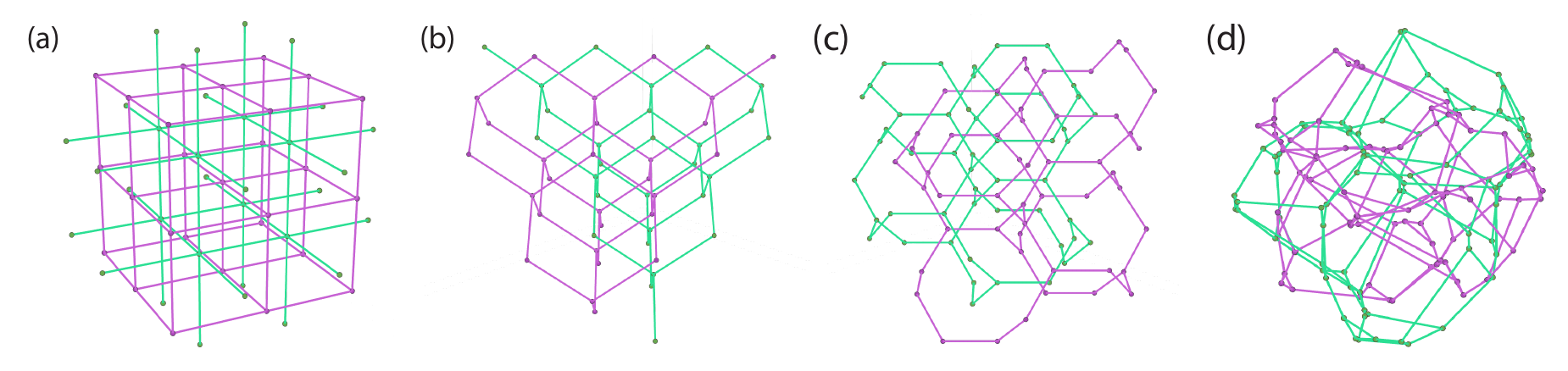} \vspace{-2mm}
    \caption{Lattice network structures. (a) A unit-cube network ensnarled with a $p=2$ cubic lattice. (b) Two ensnarled diamond networks generated with $p=2$ using translations of the underlying face-centered cubic lattice. (c) Two ensnarled Laves networks generated with $p=2$. (d) Two ensnarled isomorphic Voronoi networks with $p_s=3$ seed points.
}
    \label{fig:lattices}
\end{figure}

We consider three highly ordered three dimensional lattice systems of increasing structural complexity: the cubic, diamond, and Laves lattices. 
These networks are closely associated with the canonical triply periodic minimal surfaces, with the cubic, diamond and Laves graphs forming the skeletal networks of the Schwarz P, Schwarz D, and gyroid surfaces respectively \cite{schoen1970infinite}. 
Their well defined periodic embeddings and regular graph topologies make them convenient model systems for examining how network architecture shapes the ensnarlment spectrum
% These lattices have a well defined periodic embedding and regular graph topology, making them convenient models. 
(\figurename~\ref{fig:lattices}).

The cubic lattice is composed of vertices located at integer coordinates in space, with each vertex connected to 6 nearest neighbors with unit length edges
(\figurename~\ref{fig:lattices}a).

The diamond lattice consists of two interpenetrating FCC sublattices displaced by $\left( \frac{1}{4}, \frac{1}{4}, \frac{1}{4} \right)$
(\figurename~\ref{fig:lattices}b).
Every vertex has degree 4, with edges arranged tetrahedrally in 3 dimensions. The shortest cycles are of length 6, producing a highly regular cyclic
 structure. 
Beyond its importance as a canonical crystal lattice \cite{bonneau2015high}, the double diamond morphology has been observed in biological systems, such as mitochondrial networks formed by giant amoebae under starvation \cite{deng1998three}.

The Laves lattice forms a 3D network of edge-sharing tetrahedra generated by iteratively attaching tetrahedra along shared edges
(\figurename~\ref{fig:lattices}c).
Like the diamond lattice, it is locally tetrahedral but exhibits a substantially rich cycle structure due to the extensive sharing of edges. 
The shortest cycles are of length 10, and neighboring cycles may share one to five edges, producing a highly interconnected cycle space while preserving periodic order. 
Unlike the cubic and diamond lattices, the Laves network is inherently chiral. 
Accordingly, the double Laves systems that we consider in this work consist of two interpenetrating lattices of opposite chirality. 
Related gyroid geometries also occur in biological photonic structures, including those responsible for structural coloration in butterfly wings \cite{michielsen2007gyroid} and bird feathers \cite{saranathan2021evolution}.

To complement these lattices with a disordered network architecture, we additionally consider three-dimensional Voronoi networks.
Random seed points are distributed within a cubic domain and are periodically mirrored outside the domain to construct the boundary cells. 
A 3D Voronoi tessellation is then performed on the extended point set, after which only the cells associated with the seed points in the original cubic domain are retained. 
This produces a finite disordered network without truncated Voronoi cells at the domain boundaries. 
Two such Voronoi graphs may be generated independently to form an ensnarled system. 
In our work, we consider isomorphic ensnarled pairs, where the first network is copied and re-embedded by cyclically permuting its spatial coordinates about the geometric center: $(x,y,z) \rightarrow (z,x,y)$. 
This preserves the complete graph topology while producing a distinct spatial embedding for the second network
(\figurename~\ref{fig:lattices}d).

\begin{figure}[]
    \centering  
    \includegraphics[width=0.8\linewidth]{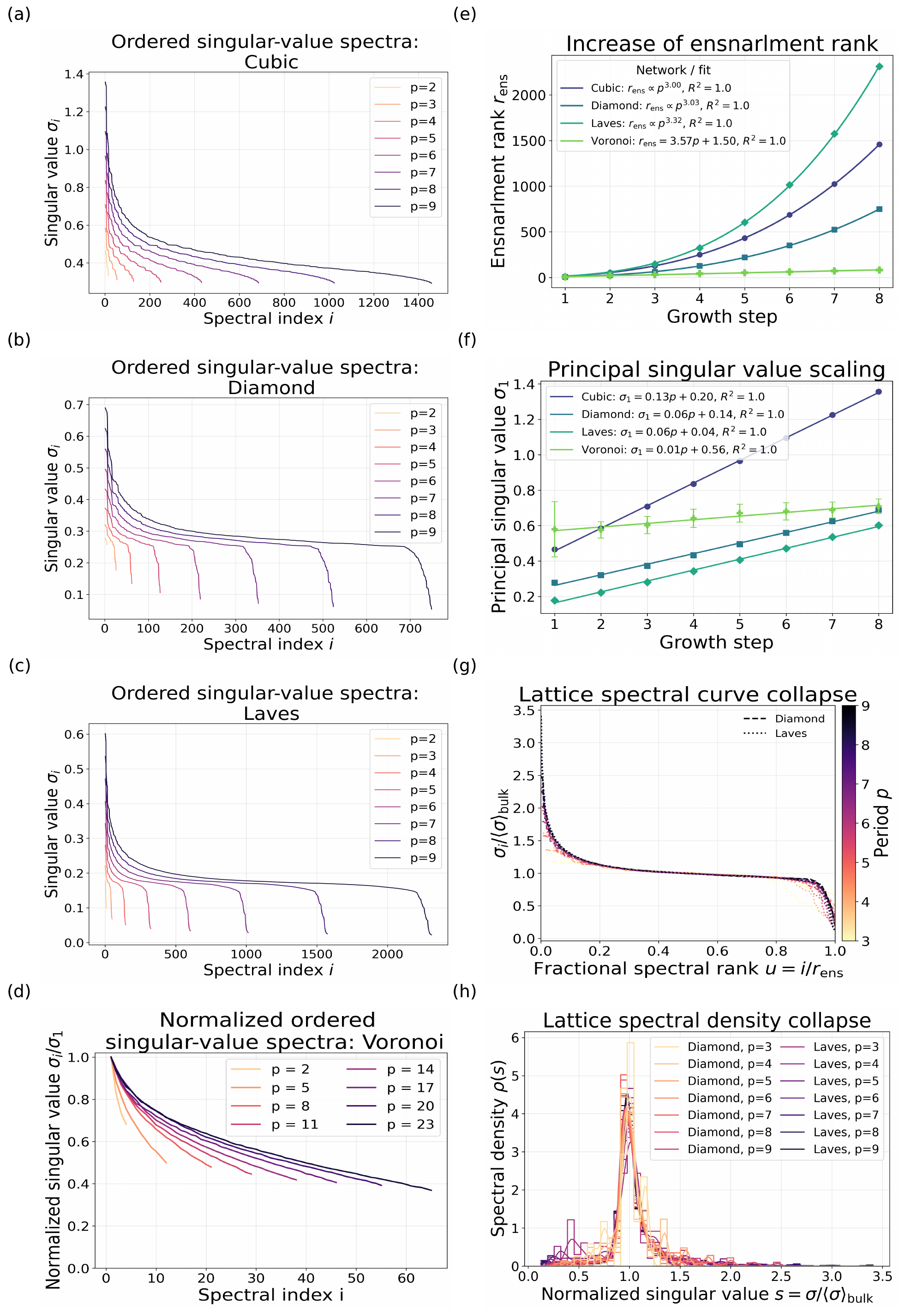}
    \caption{
        (a–d) Ordered singular value spectra for (a) cubic, (b) diamond, (c) Laves, and (d) Voronoi networks across eight increasing values of the growth parameter $p$. Nonzero singular values are ordered in descending magnitude. For Voronoi networks, each realization is normalized by its principal singular value and the pointwise median over 100 realizations is shown.
        (e) Ensnarlment rank $r_{\mathrm{ens}}=\mathrm{rank}(\Lambda)$ over eight growth steps. For ordered lattices, steps $1,\ldots,8$ correspond to $p=2,\ldots,9$; for Voronoi networks, to $p=2,5,8,\ldots,23$, with ensemble averages over 100 realizations. Fits use the underlying $p$. The ordered lattices scale approximately cubically, $r_{\mathrm{ens}}\sim p^\alpha$ with $\alpha\approx3$, whereas Voronoi networks scale approximately linearly, reflecting their respective growth constructions. 
        (f) Principal singular value $\s_\mathrm{principal}$ over the same growth steps. Linear fits $\s_\mathrm{principal} = mp+b$ are performed against $p$, showing approximately linear growth of the dominant ensnarlment mode over the sampled range, with no size-independent saturation.
        (g) Spectral curve collapse for the diamond and Laves lattices of $p\geq3$. Spectra are plotted against the fractional rank $u_i=i/r_{\mathrm{ens}}$ and normalized by the mean bulk singular value $\langle\sigma\rangle_{\mathrm{bulk}}$. The rescaled spectra collapse closely onto a common, approximately size-independent curve.
        (h) Spectral distributions for the diamond and Laves lattices of $p\geq3$, with singular values scaled as in (g). Histograms use shared bin edges and are independently normalized to unit area. Smooth curves show kernel density estimates.
        % Variability of Voronoi spectra. Raw ordered spectra from 100 realizations are shown for $p=5,11,$ and $20$, with the median over their common spectral support highlighted.
        }
    \label{fig:lattice_spectra}
\end{figure}

We now investigate how the ensnarlment spectra in these systems evolve with increasing size. For each lattice family, the size is systematically increased by replicating the unit cell along all three spatial dimensions, and the singular values of the corresponding ensnarlment operator were computed for each realization
(\figurename~\ref{fig:lattice_spectra}).

%%% Cubic increase in degree of ensnarlment %%%
% As the lattice size increases, the number of nonzero singular values increases for each of the systems. 
% Since the lattices are enlarged along the three spatial axes, the number of cycles and independent linked cycle interactions grow with the lattice volume. 
% Consequently, the ensnarlment rank exhibits an approximately cubic dependence on the lattice period $p$. 
% By contrast, the Voronoi networks are grown by increasing the number of seed points, and their ensnarlment rank grows approximately linearly 
% (\figurename~\ref{fig:lattice_spectra}e).

In all systems, the ensnarlment rank increases with network size.  
For the periodic lattices, size is controlled by repeating the unit cell $p$ times along each of the three spatial axes.  
The number of unit cells, and consequently the number of cycles and independent linked-cycle interactions, therefore grows approximately as $p^3$.  
The ensnarlment rank accordingly also exhibits an approximately cubic dependence on $p$.  
For the Voronoi networks, size is instead parameterized directly by the number of seed points, and the ensnarlment rank increases approximately linearly with this number (\figurename~\ref{fig:lattice_spectra}e).  
% \cs{ 
Both network classes therefore exhibit approximately proportional growth of the ensnarlment rank with network size, with some geometry-dependent deviations.
% The different apparent scaling laws therefore reflect the different parameters used to control network size.
% }
% \yu{[This paragraph needs reconstruction; ``By constrast'' needs rephrasing.]}

%%% Distribution and spacing 

% While the number of singular values quantifies the growth of the spectrum, it doesn't describe how the spectral modes are distributed. 
% We therefore examine the singular value distributions and their ordering across increasing system sizes. 

To study the 
%growth and 
internal organization of the full singular spectrum,
% singular value spacings and distributions, 
we examine the singular value ordering across increasing system sizes
(\figurename~\ref{fig:lattice_spectra}a-d).
The ordered spectra change with network size in both their spectral rank and characteristic singular value scale. 
% show that the singular values densely aggregate in broad regions where adjacent singular value spacings are very small. 
% This behavior is more clearly illustrated by the ordered singular value spectra, which have similar overall profiles across all sizes. 
Increasing the size primarily extends the spectrum by introducing additional singular values while preserving its shape, especially for the diamond and Laves networks. 

To compare the two networks' spectral shapes independently of rank and overall magnitude, we remove these two size dependent scales to collapse the two sets of curves
(\figurename~\ref{fig:lattice_spectra}g).
The smallest system with $p=2$ is excluded from this analysis to reduce the influence of the smallest finite-size network. 
% To compare spectral shapes independently of rank and overall magnitude, we rescale both axes. 
For the ordered nonzero singular values $\s_i$, we define
\[
    u_i = \frac{i}{r_{\mathrm{ens}}}, \qquad 
    \widetilde{\s}_i = \frac{\s_i}{\langle \s\rangle_{\mathrm{bulk}}},
\]
where the fractional rank $u_i$ maps spectra of different ranks onto $0<u<1$.
% The half-index shift places each discrete singular value at the midpoint of its fractional rank interval, rather than at an interval boundary, and avoids assigning the first and last modes to the endpoints $u=0$ and $u=1$. 
$\langle \s\rangle_{\mathrm{bulk}}$ is the mean of the central $50\%$ of the spectrum, corresponding to $0.25<u<0.75$. 
This bulk normalization removes the characteristic singular value scale without allowing the principal modes or lower spectral tail to dominate. 
% The $p=2$ systems are excluded to reduce the influence of finite-size effects.

After rescaling, the diamond and Laves spectra collapse closely onto a common curve 
(\figurename~\ref{fig:lattice_spectra}g). 
Within each lattice family, the curves are approximately independent of $p$, and the curves for the two families largely coincide despite their different network geometries. 
The spectra therefore approximately obey
\[
    \s_i^{(\mathrm{Lat},p)} \simeq 
    \langle \s\rangle_{\mathrm{bulk}}^{(\mathrm{Lat},p)} F\!\left( \frac{i} {r_{\mathrm{ens}}^{(\mathrm{Lat},p)}} \right),
\]
where $\mathrm{Lat} \in \{ \mathrm{diamond},\mathrm{Laves} \}$ denotes the lattice family \cite{bhattacharjee2001measure}. 
The normalized profile $F(u)$ is approximately independent of both lattice size and lattice type over the systems considered, indicating that growth changes the number and absolute strengths of the ensnarlment modes while approximately preserving their relative distribution.

% \textcolor{blue}{figure(s): singval distribution hist, adjacent singval spacing}

%%% Principal singval

Finally, we consider the principal singular value $\s_1$, which quantifies the maximum amplitude of the ensnarlment operator and therefore the strength of the dominant singular mode. 
For each lattice system, the principal singular value increases approximately linearly with the size of the lattice. 
No evidence of saturation is observed, over the sizes considered. 
Although these lattices are geometrically regular, extensive edge sharing among cycles produces Gram and linkage matrices with a significantly more intricate structure than the ladders or grids. 
Furthermore, each increase in $p$ adds an outer layer of ensnarled cycles to the existing network, allowing the principal mode to extend over the progressively larger coupled structure. 

\begin{figure}[h]
    \centering
    \includegraphics[width=\linewidth]{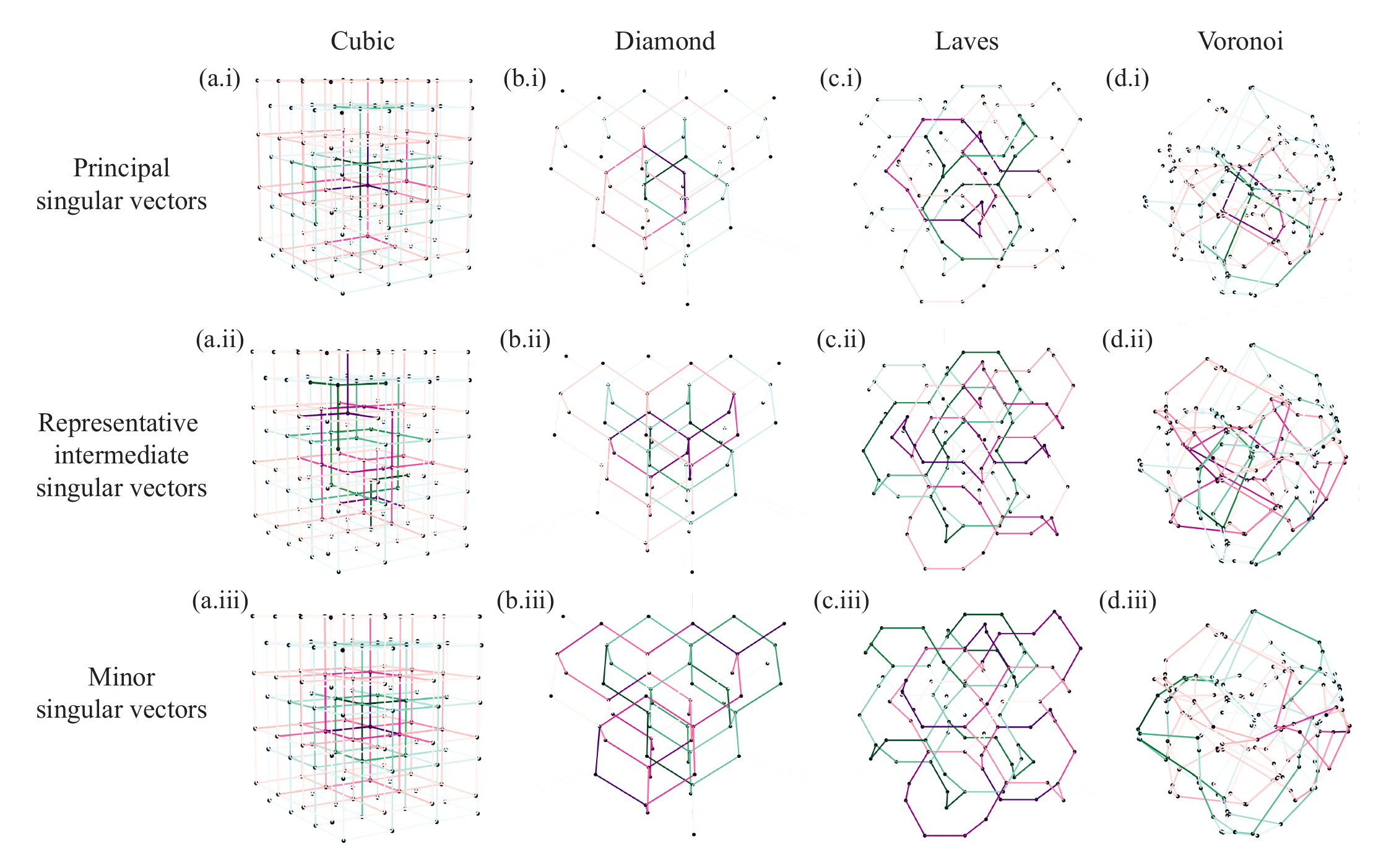}
    \caption{Singular vector pairs of the ensnarled networks shown in \figurename~\ref{fig:lattices}. Columns show the cubic, diamond, Laves, and Voronoi networks, respectively. 
    (a.i--d.i) Principal singular vectors associated with the largest singular values. 
    For degenerate singular values, the displayed vectors represent one choice of orthonormal basis within the corresponding singular subspace and are therefore not individually unique. 
    (a.ii--d.ii) Intermediate singular vectors associated with singular values near the middle of the ordered nonzero spectrum. 
    (a.iii--d.iii) Minor singular vectors associated with the smallest nonzero singular values. 
    The two color families distinguish the two networks. 
    Edge darkness encodes the magnitude of the corresponding component of the unit-normalized singular vector: darker edges have larger component magnitudes and participate more strongly in that singular mode. 
    Color intensity should therefore be interpreted relative to other edges within the same displayed vector.}
    \label{fig:singvecs}
\end{figure}

The spatial organization of the singular vectors is particularly clear in the diamond and Laves networks
(\figurename~\ref{fig:singvecs}b.i--b.iii,c.i--c.iii). 
Their principal singular subspaces identify edges in the central region of the ensnarlment, while modes associated with smaller singular values progressively redistribute their weight toward other spatial regions. 
The cubic lattice exhibits a less direct spatial ordering: although the principal mode emphasizes central edges, the minor mode also retains substantial contributions from the innermost region (\figurename~\ref{fig:singvecs}a.i--a.iii), indicating that singular value ordering does not necessarily correspond to a strict progression from central to peripheral edges. 
A similar, although less regular, organization is observed in the disordered Voronoi networks. 
Despite the absence of lattice symmetry, the principal singular vectors remain concentrated on edges occupying the central region of the ensnarlment 
(\figurename~\ref{fig:singvecs}d.i--d.iii). 
The spatial organization of the singular modes is therefore observed across both periodic and disordered systems, while its detailed form depends on the underlying network architecture.

The singular values and vectors together provide a complete modal description of the ensnarlment operator, characterizing both the strength and spatial organization of individual ensnarlment modes. 
This raises the question of how the overall complexity of an ensnarled system can be summarized by a single global quantity.

\section{Degree of Ensnarlment} \label{sec:degree_ensnarlment}

% \cs{
%     \begin{definition}[Ensnarlment centrality]
%         Let 
%         \[
%             I_\s = \{ i: \s_i = \s \}
%         \]
%         index the singular vectors associated with a singular value $\s$. 
%         The left and right ensnarlment centralities of edges $e \in E_1$ and $f \in E_2$ with respect to this singular subspace are 
%         \[
%             c^{(1)}_e(\s) = \sum_{i \in I_\s} |U_{ei}|^2,
%             \qquad
%             c^{(2)}_f(\s) = \sum_{i \in I_\s} |V_{fi}|^2
%         \]
%         For a nondegenerate singular value, these expressions reduce to the squared components of the corresponding left and right singular vectors. 
%     \end{definition}
%     This definition is analogous in a broad spectral sense to graph eigenvector centrality, in which components of the principal eigenvector assign centrality scores to nodes \cite{bonacich1972factoring}. 
%     Here, however, the scores are defined on the edges of the two ensnarled networks and are resolved by the ensnarlment mode. 
%     Particularly, ensnarlment centrality does not assume that edges with large scores occupy the geometric center of the system. 
% }

The ensnarlment rank introduced in Sec.~\ref{sec:spectral_characterization} provides a global quantity to characterize ensnarlment:
\begin{equation}
    r_{\mathrm{ens}} = \operatorname{rank}(\Lambda) = |\{i:\sigma_i>0 \}|.
\end{equation}
Thus, $r_{\mathrm{ens}}$ counts the nonzero singular modes supported by the coupled network pair. 
We interpret this global mode count as the \textit{degree of ensnarlment}. 

The structural meaning of this quantity follows from the relation between $\L$ and the linkage matrix $L$.
Since the mesh matrices $M_i$ have full column rank and the Gram matrices $G_i$ are invertible, multiplication by the corresponding edge space embedding factors preserves rank. 
Consequently,
\begin{equation}
    r_{\mathrm{ens}} = \operatorname{rank}(\L) = \operatorname{rank}(L).
\end{equation}
Each nonzero singular mode of $\L$ therefore represents an independent direction in linkage space, corresponding to an independent linkage interaction.
% or combination of interactions between the two cycle spaces.
% This quantity admits a natural topological interpretation. 
% Each nonzero singular mode represents an independent mode of interaction between linked cycle pairs within the two networks. 
The degree of ensnarlment therefore measures the number of independent linked interactions present in the system. 
Unlike the individual singular values which quantify the strength of particular interactions, it summarizes the overall complexity of the coupled network in a single scalar quantity.

The geometric interpretation of these independent modes follows directly from the singular vectors discussed in Sec.~\ref{sec:spectral_properties}. 
In structured networks, each singular mode corresponds to a distinct spatial organization of edges, producing edge patterns that extend from the most central regions of the ensnarled system towards its periphery. 
The degree of ensnarlment therefore counts the number of independent spectral modes required to represent the complete ensnarled structure.

To exemplify this interpretation, we compare the ensnarlment rank across several classes of networks. 
Minimally linked systems exhibit a single independent interaction and therefore have $r_{\mathrm{ens}}=1$. 
As network complexity increases through shared-edge ladder networks and periodic lattice systems, additional independent interaction modes emerge, leading to a corresponding increase in the ensnarlment rank. 
Highly symmetric lattices further exhibit degenerate singular modes arising from equivalent symmetry directions, while irregular Voronoi networks produce distinct nondegenerate modes reflecting the absence of global symmetry.
% \cs{sounds like i planned a figure here}

Beyond providing a global measure of ensnarlment, the ensnarlment rank also establishes a direct connection to the problem of separating two coupled networks. 
As shown in the following section, the ensnarlment rank bounds the minimum number of edge removals required to completely eliminate the ensnarlment, providing both a quantitative measure of topological complexity and the theoretical basis for a spectral edge cutting algorithm.

\section{Edge cutting algorithm} \label{sec:edge_cutting}

Kramer and Modes introduced a greedy cutting procedure in which one network is designated as the target and the other as the reference \cite{kramer2023ensnarled}. 
At each iteration, each target edge is assigned a priority equal to the squared norm of its corresponding row or column of $\L$, depending on which network is chosen as the target. 
The edge with the largest priority is removed, the linkage matrix $L$ and ensnarlment operator $\L$ are recomputed, and the procedure continues until $L$ vanishes. 
Although this method identifies efficient cut sets in many examples, it collapses the complete singular structure into a single aggregate priority for each edge and greedily selects the current maximum. 
It can therefore favor peripheral or discretization-dependent edges and does not ensure that every removal eliminates an independent linkage mode or contributes to a globally minimal cut set. 

Here, we instead use the resolved singular subspaces of $\L$ to identify edges participating in individual ensnarlment modes and evaluate each accepted removal through the resulting reduction in ensnarlment rank. 
This directly connects the cutting procedure to the elimination of independent linkage modes, treats degenerate singular subspaces consistently, and provides the rank-based lower bound against which the resulting cut set size can be assessed.

The preceding greedy construction raises a more fundamental question: what is the minimum number of edge removals required to eliminate the ensnarlment completely? 
Because removing a single edge can eliminate at most one independent direction from the cycle space of the target network, it can reduce $\operatorname{rank}(L)=\operatorname{rank}(\L)$ by at most one. 
Complete unlinking requires this rank to be reduced from $r_{\mathrm{ens}}$ to zero, giving the following lower bound.

\begin{theorem}[Rank lower bound for unlinking]\label{thm:rank_lower_bound}
    Let $\X\subseteq \mathcal{E}_{\mathrm{target}}$ be any set of target network edges whose removal eliminates all linkage between the two networks. 
    Then
    \begin{equation}
        |\X|\geq r_{\mathrm{ens}}.
    \end{equation}
    In particular, the minimum unlinking cut set satisfies
    \begin{equation}
        |\X_{\min}|\geq r_{\mathrm{ens}}.
    \end{equation}
\end{theorem}

The proof is given in Appendix~\ref{app:rank_lower_bound}. 
This bound provides a natural benchmark for evaluating candidate cut sets. 
We next introduce a cutting algorithm based on the dominant singular subspaces of $\L$ and evaluate it across ordered lattice and irregular Voronoi networks. 
% For the ordered lattices, the resulting cut sets attain the rank lower bound, whereas the Voronoi networks exhibit non-rank-tight behavior and provide a more stringent test of the algorithm.

% While the ensnarlment rank provides a quantitative measure of the complexity of an ensnarled system, it does not indicate how this complexity may be reduced. 
% This naturally leads to the  the problem of unlinking two ensnarled networks through edge removal. 
% Specifically, we seek the smallest set of edge deletions required to eliminate all linked cycle interactions, thereby reducing the ensnarlment rank to zero. 

% In this section, we first establish that the minimum edge cut is bounded below by the ensnarlment rank, demonstrating a fundamental connection between the spectral properties of $\L$ and the topology of the coupled network pair. 
% We then introduce an edge cutting algorithm based on the leading singular subspace fo the ensnarlment operator. 
% The algorithm is evaluated on a range of network classes, where we show that the leading eigenspace is sufficient to recover the minimum cut set in highly ordered lattice systems, before examining its robustness in irregular Voronoi networks. 

\subsection{Edge cutting algorithm}

% The rank lower bound establishes the smallest number of edge removals that could, in principle,completely eliminate the ensnarlment. 
% It does not, however, identify which edges should be removed. 
To construct candidate edge cut sets, we use the singular subspaces of the original ensnarlment operator to score edges according to their participation in the corresponding ensnarlment modes.

Let
\begin{equation}
\L = U \S V^T
\end{equation}
be the SVD of the original operator, where the columns of $U$ and $V$ define singular vectors on the edge spaces of the first and second networks, respectively. 
The SVD is computed once before any edges are removed, and the resulting singular subspaces and edge scores remain fixed throughout the cutting procedure.

We set one network as the target network from which edges are cut. 
Suppose that the first network is chosen.
For a nonzero singular value $\s_\a$ of multiplicity $m_\a$, let
\[
U_\a =
\begin{pmatrix}
    \bm{u}_{\a,1} & \cdots & \bm{u}_{\a,m_\a}
\end{pmatrix}
\]
span its left singular subspace.
Because the basis within a degenerate singular subspace is not unique, edges are scored using their total squared projection onto the complete subspace:
\begin{equation}
    s^{(1)}\a(e) = \sum_{i=1}^{m_\a} \left| \bm{u}_{\a,i}(e) \right|^2.
\end{equation}
This score is invariant under orthogonal changes of basis within the singular subspace and is equivalent to the ensnarlment centrality introduced in Sec.~\ref{sec:spectral_characterization}.  
If the second network is chosen as the target, the corresponding scores are obtained from the right singular subspaces of $V$.

The distinct nonzero singular subspaces are considered in decreasing order of their singular values. Within each selected subspace, uncut edges with scores above a numerical threshold are tested as candidate cuts.
At each iteration, all eligible single edge cuts are evaluated, and the candidate producing the largest reduction in residual rank is selected.
The singular subspace score and residual linkage norm are used successively to distinguish candidates with equal rank reduction.
If no single edge reduces the rank, pairs of eligible edges are tested in the same manner.
When the current singular subspace cannot reduce the rank further, the procedure continues to the next selected subspace.

The effect of a provisional cut set is evaluated by restricting the cycle space of the target network.
For a provisional cut set $\X$, let $M_1[\X,:]$ denote the rows of the mesh matrix $M_1$ corresponding to the cut edges, and let $Q_\X$ be an orthonormal basis for
\begin{equation}
    \ker \! \left( M_1[\X,:] \right).
\end{equation}
The columns of $Q_\X$ represent the cycle combinations having zero amplitude on every edge in $\X$.
The residual mesh and linkage matrices are therefore
\begin{equation}
    M_{1,\X}=M_1Q_\X, \qquad L_\X=Q_\X^TL.
\end{equation}
The corresponding residual ensnarlment operator is
\begin{equation} 
    \L_\X = M_{1,\X}^{+^T} \, L_\X M_2^{+},
\end{equation}
where $+$ denotes the generalized inverse. 
This construction obtains the residual linkage directly from the original cycle linkage matrix and therefore avoids recomputing the Gauss linking numbers for every provisional cut.

Since the residual mesh matrices retain full column rank,
\begin{equation}
    \operatorname{rank}(\L_\X) = \operatorname{rank}(L_\X).
\end{equation}
At iteration $i$, a candidate edge is accepted only if it reduces the residual rank, satisfying
% the resulting cut set $\mathcal{X}_i$ satisfies
\begin{equation}
    \operatorname{rank}\left(L_{\mathcal{X}_i}\right)
    <
    \operatorname{rank}\left(L_{\mathcal{X}_{i-1}}\right).
\end{equation}
Otherwise, the edge is restored before the next candidate is evaluated.
When no rank reducing single edge remains, the same criterion is applied to candidate edge pairs.

This procedure distinguishes spectral information used to rank edges from the rank test used to accept cuts. 
The original SVD identifies edges that participate in the most strongly ensnarled modes, whereas rank of the residual operator determines whether a proposed removal eliminates an independent linked cycle interaction. 
Consequently, high spectral score alone is not sufficient for accepting a cut; an edge is retained unless its removal produces an actual reduction in the ensnarlment rank. 

The algorithm can be used to determine whether the rank lower bound is attained for a given ensnarled system.
If it produces an unlinking cut set $\X$ with $|\X|=r_{\mathrm{ens}}$, then the lower bound proves that no smaller cut set exists. 
We refer to such systems as \textit{rank-tight}. 

\begin{definition}[Rank-tight ensnarlment]
    A system is rank-tight if $|\X_\mathrm{min}| = r_\mathrm{ens}$.
\end{definition}

We apply the algorithm to the network classes considered above.
For the highly ordered diamond and Laves lattices, it identifies cut sets whose cardinalities equal the ensnarlment rank, thereby establishing that these systems are rank-tight.
For the Voronoi networks, by contrast, the cut sets returned by the fixed-SVD algorithm contain more than $r_{\mathrm{ens}}$ edges.
This observed cut excess may arise because the singular subspaces, computed only once from the original operator, become progressively less representative of the residual system after successive edge removals. 

To test this possibility, we introduce an iterative variant in which the residual ensnarlment operator and its SVD are recomputed after every accepted cut.
Edge scores are then obtained from the updated principal singular subspace.
All remaining edges are retained as candidates, including those with zero participation in this subspace, so the spectral scores determine the order and priority of the search rather than its admissible edge set.
The rank reduction criterion used to accept cuts remains unchanged.

\begin{figure}[h]
    \centering
    \includegraphics[width=\linewidth]{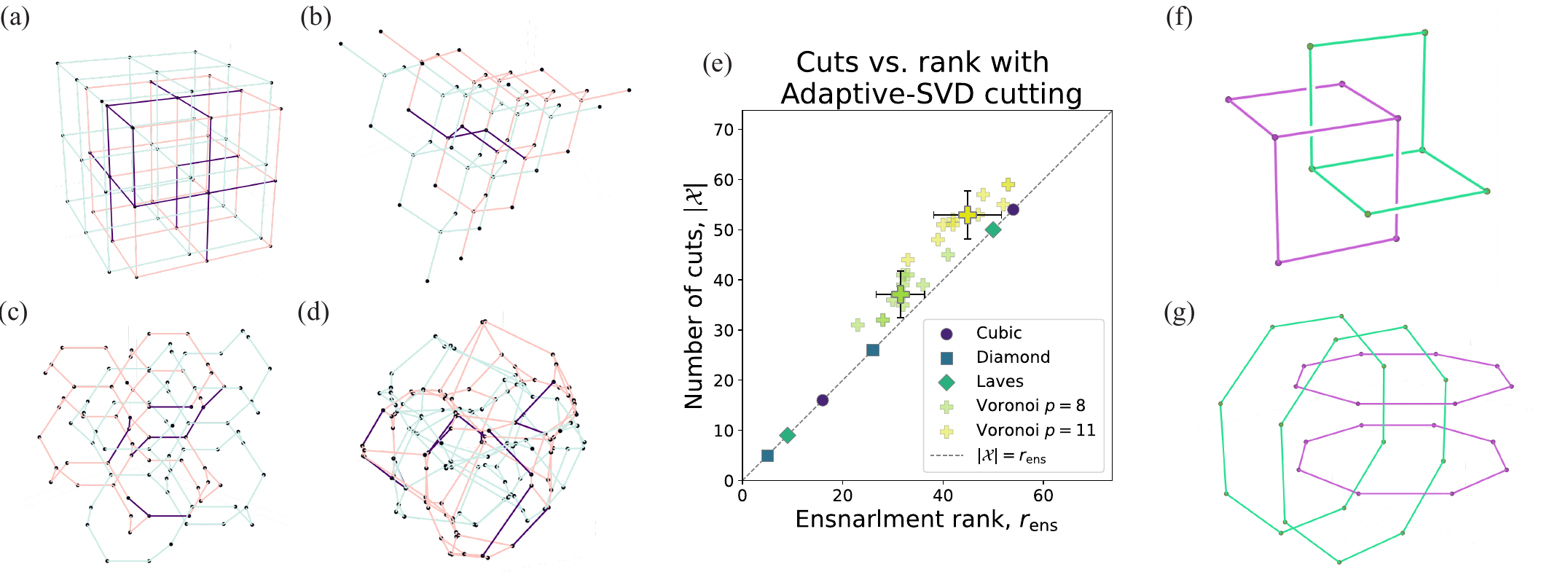}
    \caption{(a-d) Spectral edge cuts in representative 3D ensnarled networks: (a) cubic, (b) diamond, (c) Laves, and (d) Voronoi. The displayed networks are the same representative systems shown previously, except for the cubic case, which consists of two ensnarled cubic lattices with length $=2$. In each panel, the purple network is chosen as the target for cutting; cut edges are highlighted by the darker color, while uncut edges are shown in the corresponding lighter shade. 
    % (e) Difference in cut set size obtained by recomputing the SVD after each cut, shown for 10 independent Voronoi realization at periods $p = 8, 11$. 
    (e) Rank tightness across the four 3d network classes. Two system sizes are shown for each class: $p=2,3$ for the cubic, diamond, and Laves lattices, and $p=8,11$ for the Voronoi networks. The ordered lattices are rank-tight and therefore lie exactly on the $y=x$ line, whereas the Voronoi realization exhibit non-rank-tight behavior. Individual Voronoi realizations are shown with reduced opacity, while the mean cut set size versus the mean ensnarlment rank is highlighted with full opacity. None of the Voronoi realizations shown are rank-tight. 
    (f) Minimal example of rank tightness: ladders with $|\X_\mathrm{min}| = r_\mathrm{ens} = 2$.
    (g) Minimal example of cut excess: a double Hopf-link configuration with $\Delta_\mathrm{cut} = 1$.}
    \label{fig:edge_cuts}
\end{figure}

For the ordered lattice systems, the adaptive-SVD algorithm again produces cut sets whose cardinalities equal the ensnarlment rank.
For the Voronoi networks, the fixed and adaptive approaches produce similar cut numbers
(\figurename~\ref{fig:edge_cuts}f).
Across the 10 sampled realizations, recomputing the SVD either leaves the cut number unchanged or alters it by a single edge, with a tendency toward fewer cuts under recomputation.
Nevertheless, both algorithms return cut sets larger than the ensnarlment rank
(\figurename~\ref{fig:edge_cuts}e).

The persistence of this cut excess under adaptive-SVD recomputation indicates that the use of a fixed singular vector ranking is not its sole cause.
However, because neither spectral procedure performs an exhaustive optimization over all possible cut sets, these results establish only that rank-tight cut sets are not recovered for the Voronoi networks by either of the heuristics considered here.
They do not determine whether the Voronoi networks are intrinsically non-rank-tight.
Intrinsic non-rank-tightness is nevertheless possible, as demonstrated by the following minimal example.

Consider a double Hopf-link where two networks are composed of two edge-disjoint cycles 
(\figurename~\ref{fig:edge_cuts}g) with linkage matrix
\[
    L = \begin{pmatrix}
        -1 & 1 \\ 1 & -1
    \end{pmatrix}.
\]

Although $\operatorname{rank}(L) = 1$, removing a single edge cannot eliminate both linked interactions simultaneously. 
Consequently, any complete unlinking requires two edge removals, despite the spectral lower bound predicting only one. 
% Thus, multiple edge removals are required before the ensnarlment rank decreases by one.
% Hence,
% \[
%     |X_{\mathrm{min}}|=2>r_{\mathrm{ens}}=1.
% \]
The rank therefore provides a lower bound of one cut but does not capture the additional removal imposed by the edge-disjoint realization of the linked cycles.

This motivates a quantitative measure of the deviation from rank-tightness:
% We therefore define \textit{cut excess} as the number of edges required beyond the spectral rank bound,
\begin{definition}[Cut excess]
The \emph{cut excess} of an ensnarled system is defined as
\[
    \Delta_{\mathrm{cut}} = |\X_{\min}|-r_{\mathrm{ens}},
\]
where $|\X_{\mathrm{min}}|$ is the cardinality of a minimum edge cut set.
A rank-tight system satisfies $\Delta_{\mathrm{cut}}=0$, whereas $\Delta_{\mathrm{cut}}>0$ indicates that additional edge removals beyond the spectral lower bound are required for complete unlinking.
\end{definition}

For the double-Hopf example shown above, $\Delta_{\mathrm{cut}}=1$.
When the minimum cut is not known exactly, as for the Voronoi networks considered here, an algorithmic cut set $X_{\mathrm{alg}}$ instead defines the observed cut excess
\begin{equation}
    \widehat{\Delta}_{\mathrm{cut}} = |\X_{\mathrm{alg}}|-r_{\mathrm{ens}}.
\end{equation}

Since $\X_{\mathrm{alg}}$ is a feasible but not necessarily a minimum cut set, 
\begin{equation}
    0 \leq \Delta_{\mathrm{cut}} \leq \widehat{\Delta}_{\mathrm{cut}}.
\end{equation}

Thus, a positive observed cut excess quantifies the failure of a particular algorithm to attain the  rank bound but does not prove that the system is intrinsically non-rank-tight.

\section{Graph Laplacian view}
\label{sec:laplacian}
% \textcolor{magenta}{unbalance score from the laplacian representation}
In this section, we discuss in more detail the ensnarlment operator $\L$ from the graph Laplacian view. 

As discussed in Sec.~\ref{sec:theoretical_overview}, the entries of $\L$ encode interactions between the edges induced by linked cycles, and may be interpreted as the weights of a graph $\mathcal{G}$ with vertex set $\mathcal{E}_1 \cup \mathcal{E}_2$.
These vertices correspond to the edges of the two original networks are are hereafter called spatial edges. 
% a weighted adjacency matrix that describes a graph structure, {$\mathcal{G}$}, between the two sets of edges $\mathcal{E}_1, \mathcal{E}_2$ (spatial edges hereafter). 
The induced connections only exist between the two sets $\mathcal{E}_1$ and $\mathcal{E}_2$ but not within them; therefore, $\mathcal{G}$ is bipartite. 
Furthermore, since linking numbers are signed, the entries of $\L$ can be positive or negative, making $\mathcal{G}$ a signed graph. 
% and thus $\L$ can take positive or negative values; therefore, $\mathcal{G}$ is also signed. 

To summarize, $\L$ induces a signed, weighted bipartite graph $\mathcal{G} = (\mathcal{V}_b, \mathcal{E}_b)$, where
\[
    \mathcal{V}_b = \mathcal{E}_1 \cup \mathcal{E}_2,
    \qquad
    \mathcal{E}_b = \left\{ (e,f) \in \mathcal{E}_1 \times \mathcal{E}_2 : \L_{ef} \neq 0 \right\}.
\]
Its signed weighted adjacency matrix is
\begin{align*}
    % \mathbf{W}_b =
    W_b =
    \begin{pmatrix}
        \mathbf{0} & \L \\
        \L^T & \mathbf{0}
    \end{pmatrix}
    = 
    \begin{pmatrix}
        \mathbf{0} & {M_1^+}^T \cdot L \cdot M_2^+ \\
        {M_2^+}^T \cdot L^T \cdot M_1^+ & \mathbf{0}
    \end{pmatrix}= 
      \begin{pmatrix}
        M_1^+ & \mathbf{0} \\
        \mathbf{0} & M_2^+
    \end{pmatrix}^T
     \begin{pmatrix}
        \mathbf{0} & L \\
        L^T & \mathbf{0}
    \end{pmatrix}
     \begin{pmatrix}
        M_1^+ & \mathbf{0} \\
        \mathbf{0} & M_2^+
    \end{pmatrix}
    % \\
    % &= \begin{pmatrix}
    %     M_1 & \mathbf{0} \\
    %     \mathbf{0} & M_2
    % \end{pmatrix}^{+T}
    %  \begin{pmatrix}
    %     \mathbf{0} & L \\
    %     L^T & \mathbf{0}
    % \end{pmatrix}
    %  \begin{pmatrix}
    %     M_1 & \mathbf{0} \\
    %     \mathbf{0} & M_2
    % \end{pmatrix}^+
    \, .
\end{align*}

We know that the values in the mesh matrices $M_1$ and $M_2$ depend on the orientation of edges. A natural question is then: \textit{what if we flip the orientation of a spatial edge?} Without loss of generality, flip the orientation of a spatial edge $e_i\in \mathcal{E}_1$, corresponding to the $i$-th row of the mesh matrix $M_1$. Then the mesh matrix after the orientation flip, denoted by $\tilde{M}_1$, has the element-wise opposite sign for its $i$-th row, which can be written as 
\begin{align*}
    \tilde{M}_1 = D_1M_1 \, ,
\end{align*}
where $D_1$ is a diagonal matrix with the components of $\mathbf{1} - 2\mathbf{e}_i$ on its diagonal, and where $\mathbf{1}$ is the all-ones vector, and $\mathbf{e}_i$ is the $i$-th column of the identity matrix. Since $D_1$ is symmetric and orthogonal, the pseudo-inverse of $\tilde{M}_1$ is
\begin{align*}
    \tilde{M}_1^+ = (D_1M_1)^+ = M_1^+D_1 \, .
\end{align*}
Therefore, the new ensnarlment operator $\tilde{\L}$ is
\begin{align*}
    \tilde{\L} = (M_1^+D_1)^T L M_2^+ = D_1 {M_1^+}^T  L  M_2^+ = D_1\L\, ,
\end{align*}
which also has the opposite sign for its $i$-th row. By an argument analogous to those in Lemma A.13 and Theorem A.14 in \cite{tsm2026gemini}, we obtain that the eigenvalues of the signed weighted adjacency matrix and singular values of the ensnarlment operator, remain invariant under the reversal of any subset of spatial edge orientations.
These spectral quantities therefore constitute well-defined, orientation-independent invariants of the spatial networks.

% It may therefore be possible that the spatial edges could be arranged in an aligned manner such that their pairwise ensnarlment contributions are all positive.
A natural question is whether the spatial edges can be consistently oriented such that all pairwise ensnarlment contributions are nonnegative, yielding a configuration without conflicting interactions. The existence of such an orientation is equivalent to \textit{structural balance}, a fundamental notion in the study of signed graphs \cite{harary_1953_balance}. 
The state of being structurally balanced imposes strong conditions on the relative alignment between the spatial edges. 
The deviation from structural balance therefore provides a natural measure of the system's \textit{frustration}, also called sign complexity. 
% Whether $\mathcal{G}$ is structurally balanced, and if not, how far the spatial edges can be oriented to approach structural balance, provides a natural measure to quantify the \textit{frustration} (a.k.a.~sign complexity) of the underlying system. 
% For instance, the ensnarlment operator can be interpreted within the framework of spin systems by associating each spatial edge with a spin variable and treating their entries as coupling strengths between spins. In this analogy, structural balance corresponds to the existence of a spin configuration in which all interactions are satisfied, resulting in a frustration-free state. 
The degree of frustration is commonly used to quantify \textit{unbalance} in signed networks, i.e., the opposite of structural balance \cite{harary1959frustration}. 
We propose to quantify the level of frustration induced by the ensnarlment operator through the following unbalance score:
\begin{align*}
    \l_{\min} = \min \left\{ \l : L_{\mathrm{rw}} \, \bm{x} = \l \bm{x}, \ \bm{x} \neq \bm{0} \right\},
\end{align*}
i.e., we use the smallest eigenvalue $\l_{\min}$ of the ``signed random-walk Laplacian'' 
\[
    L_{\mathrm{rw}} = I - D_b^{-1} W_b,
\]
where $D_b = \operatorname{diag}(\bm{d})$, $\bm{d}=(d_i)$, the degree $d_i = \sum_j \left| (W_b)_{ij} \right|$, $I$ is the identity matrix, and $W_b$ is the signed weighted adjacency matrix before (cf. \cite{tian2024sign}). 
Specifically, $\l_{\min} = 0$ if and only if the signed graph is structurally balanced. 
Note that there are many other walk-based measures proposed to measure the level of balance of the signed graph, e.g., \cite{estrada2014walk,kirkly2019index,singh2017index,tian2025index}, 
and we choose the eigenvalue-based one in order to preserve invariance under edge orientation. 

\begin{figure}[htbp]
    \centering
    \includegraphics[width=0.7\textwidth]{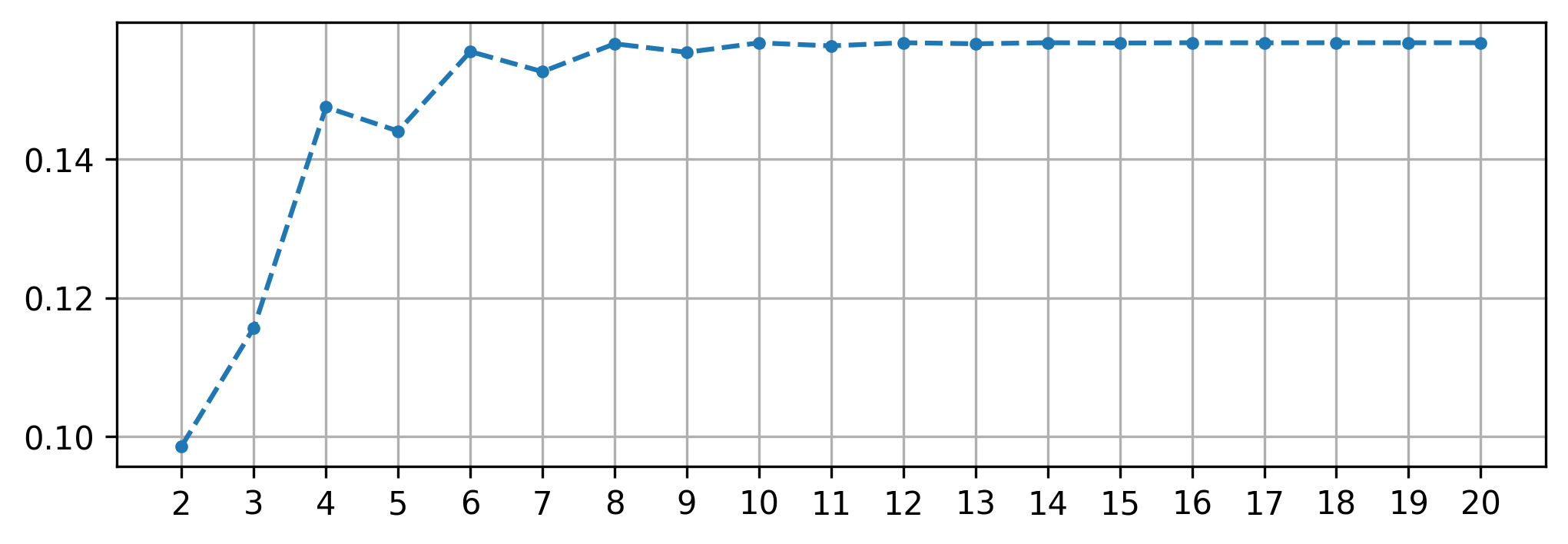}
    \caption{Unbalance scores of ensnarled ladders of increasing number of cycles (x-axis).}
    \label{fig:lad-laplacian}
\end{figure}

We examine the unbalance score for the ensnarled ladders introduced in Sec.~\ref{sec: ladders}.
As the number of cycles increases, the unbalance score exhibits an increasing trend and then stabilizes (Fig.~\ref{fig:lad-laplacian}). 
The intrinsic frustration arises because each rung of one ladder participates in two cycles but can align its orientation with only one corresponding cycle of the other ladder. 
% originates from the fact that the rungs of one ladder participate in two cycles of that ladder, but may only align their orientations with one cycle of the other. 
% the shared edges between two adjacent smallest basis cycles can only align their direction with one cycle in the other ladder that is ensnarled with one of the two basis cycles. 
Therefore, the contributions associated with the remaining unaligned cycle are necessarily negative.
% three relatively strong contributions are necessarily negative. 
As cycles are added, we thus expect the unbalance score to increase. 
% We should also note that when the number of cycles is even, the ensnarled ladders exhibit more symmetries than the odd case, which can contribute to the local downward trend as the number of cycles goes from odd to even.
However, since the unbalance score measures frustration relative to the overall interaction strength, both the frustrated and total contributions grow with system size, causing the score to stabilize. 
% in a relative sense, we also expect it to stabilize as we add more cycles systematically
Each additional cycle introduces a new shared edge with $3$ new relatively strong negative contributions, among $15$ new relatively strong contributions in total. 
This gives a heuristic saturation value of $3/15 = 0.2$. 
% In the case of ensnarled ladders, upon adding one more cycle, the new shared edge gives $3$ new relatively strong negative contributions, while providing $15$ new relatively strong contributions in total; this argument predicts that the unbalance score should stabilize at approximately $0.2$. 
We note that this prediction is an upper bound, since it does not account for additional weaker contributions that do not generate frustration.
% since many additional smaller contributions that do not contribute to frustration are not accounted for.
Further analysis of the unbalance score and the associated question of finding optimal spatial edge orientations that minimize the frustration are interesting directions to explore in the future.

\section{Discussion and Conclusion}

In this work, we investigated the spectral properties of the ensnarlment operator introduced by Kramer et al. \cite{kramer2023ensnarled} and established a spectral framework for characterizing the topology of intertwined spatial networks. 
We showed that each singular mode corresponds to an independent linked cycle interaction, with the associated singular value determined by the lengths of the participating cycles, the number of shared edges, and the cumulative contributions of all linked cycle interactions involving that mode. 
The corresponding singular vectors resolve the spatial organization of these interactions, leading to the notion of \textit{ensnarlment centrality}: an edge based analog of classical eigenvector centrality for an ensnarled network pair.

Based on this modal description, we introduced the \textit{degree of ensnarlment} as the rank of the ensnarlment operator, as a global measure of the complexity of an ensnarled system. 
This quantity counts the number of independent ensnarlment modes and provides a lower bound on the minimum number of edge removals required to completely unlink two networks. 

Exploiting the structure of the principal singular subspace, we further developed a spectral edge cutting algorithm capable of constructing complete unlinking cut sets. 
For several lattice families, the algorithm achieves the theoretical lower bound, which we characterize through the notion of \textit{rank-tightness}, while more complex systems motivate the complementary concept of \textit{cut excess} as a measure of the deviation from this optimum. 

% \yu{insert concussion from graph laplacian pov} \cs{either here or in the end? everything after this is discussion related to spectra and edge cutting} \yu{[Yu: I added two sentences at the end for us to use :-)]}

% \textcolor{blue}{Disclusion - thoughts (not a draft yet so if you're reading this, prepare to be concussed)}

% While we derive the exact singval spectra for some minimal cases, it is an increasingly difficult problem when we consider larger systems. Of course, symmetric lattices provide some order, but it is still not enough order to simplify the final closed form for the spectra. we do note that the symmetric ladders used the toeplitz tridiagonal spectra, which is characteristic to its architecture. for other larger systems where the mesh matrices or the linkage matrix cannot be reduced, determining the spectral patterns are almost impossible. however, it is possible to compute the rank of the linkage matrix and therefore establish the degree of ensnarlment, as we show in the voronoi networks. 

While exact singular value spectra can be derived for minimal network motifs, their analytical determination becomes increasingly difficult for larger systems. 
In highly structured cases, the network architecture may induce matrix forms with known spectral properties; 
for example, the symmetric ladder construction reduces to a traditional Toeplitz problem, allowing its singular spectrum to be obtained analytically. 
Such reductions are generally unavailable for larger 3-dimensional lattices and disordered systems, where neither the mesh matrices nor the linkage matrix admit a sufficiently simple form. 
Nonetheless, a complete analytical description of the spectrum is not required to determine the degree of ensnarlment. 
Since the rank of the ensnarlment operator is determined by the rank of the linkage matrix, $r_{\mathrm{ens}}$ remains directly computable even for systems whose individual singular values cannot be obtained analytically, as shown for the disordered Voronoi networks. 
% \cs{insert something about curve collapse here. interesting point: diamond and laves collapse nicely and also have same principal singval scaling}

% \cs{Boundary and symmetry effects of ensnarlment centrality} 
While ensnarlment centrality resolves the participation of individual edges in the principal singular mode, it should not be interpreted as identifying a geometric center of the system. 
The spatial distribution of edge contributions is constrained by the symmetries and boundary conditions of the networks. 
The apparent central-to-peripheral organization observed in the finite lattices should therefore be interpreted with these effects in mind. 
This is illustrated in the circular ladders considered in Appendix~\ref{app:c}, obtained by identifying the ends of an open ladder system to impose periodicity along its longitudinal direction. 
Removing this boundary eliminates any distinguished longitudinal position, and the edge contributions no longer exhibit a central-to-peripheral hierarchy, instead reflecting the periodic symmetry of the network.
A complete removal of boundary effects would require extending the present finite network framework to fully periodic or infinite ensnarled networks, which is beyond the scope of this work. 
% \cs{write the closed circular ladders into appendix}

% our edge cutting algorithm relies on the input of setting one of the networks as the target for the cuts. it is entirely possible that a smaller cut set exists if the other network is chosen as the target. furthermore, when no single edge in network 1 can reduce the rank, maybe an edge in network 2 can reduce the rank by 1, leading to a smaller cut set. 
% however, if our test system of choice happens to be ensnarled vascular and secretory flow networks, such as sinusoids and canaliculi in the liver, this feature of setting a target network might be more useful than brute-forcing a true minimum cut set. 

A further consideration arises from the construction of the edge cutting algorithm. 
In its present form, one of the two networks is designated as the target network, and all edge removals are restricted to that network. 
The resulting cut set therefore need not be the smallest possible cut set if removals from either network are permitted. 
A smaller solution may exist in the second network, or may require cuts distributed between both networks. 
In particular, a state for which no single edge in the chosen target network reduces the ensnarlment rank may still admit a rank reducing cut in the other network. 
Extending the algorithm to search jointly over both edge spaces would therefore provide a natural route towards determining globally minimal cut sets. 

The restriction to a prescribed network may, however, be advantageous in applications where modifications to the two networks are not equivalent. 
A motivating example is provided by intertwined vascular and secretory transport networks, such as the sinusoidal and bile canalicular networks of the liver. 
In such systems, the the relevant question may not be the unrestricted minimum number of cuts, but rather how the ensnarlment can be altered through interventions confined to one of the constituent networks. 
The target network formulation naturally accommodates such asymmetric constraints. 

% the notion of cut-excess demonstrates that our framework for characterizing ensnarlment falls back when there are multiple redundancies in the system, as shown in the double hopf-linked loops. these redundancies appear in the linkage matrix L as linearly dependent rows/columns, thereby rendering the matrix rank-deficient. In these cases, it is unclear if the leading SVD based cutting algorithm provides the true minimum cut set. this warrants further exploration to determine systems with different cut excess numbers.

The occurrence of nonzero cut excess exposes a separate distinction between the algebraic complexity ensnarlment and the number of physical edge removals required to eliminate it. 
Redundant cycle interactions can produce linear dependencies among the rows or columns fo the linkage matrix, reducing the rank without a corresponding reduction in the number of edges that must be removed to completely unlink the two networks. 
The double Hopf-linked system shown provides a minimal example of this behavior. 
Consequently, while $r_\mathrm{ens}$ provides a universal lower bound on the cut number, it does not in general determine it. 
Moreover, for systems with nonzero cut excess, it remains unclear whether the SVD-based cutting procedure necessarily identifies a minimum cut set. 
Characterizing the network structures that generate cut excess, determining how this excess relates to dependencies within the linkage matrix, and establishing tighter bounds on the minimum cut number therefore remain open problems. 

More broadly, our results show that the ensnarlment operator encodes substantially more structure than a collection of pairwise Gauss linking numbers. 
Its singular spectrum decomposes the interaction between two networks into distinct ensnarlment modes, its singular vectors spatially resolve these modes over constituent edges, and its rank quantifies their number. 
This spectral structure in turn provides a means of identifying the edges through which the ensnarlment can be altered. 
We also provide further analysis through the view of the graph Laplacian, paving the way towards more graph-theoretic, or mathematical in general, insights. Meanwhile, since the graph Laplacian plays a central role in various algorithms, such as graph clustering and signal processing, %\cite{}, 
this also opens up opportunities for real-world applications.

% let me throw this in too: maybe someone can use this to invent a sustainable ensnarled material. the ensnarlment will provide structural integrity and sturdiness, and breaking the ensnarlment could give two completely degradable materials. idk bro i think of the environment these days 

% \textcolor{blue}{I did warn you}

\section*{Acknowledgments}
We thank Stephan Grill and Freya Jensen for their helpful comments on this work. This work was supported by the Max Planck Society. YT acknowledges support from ELBE postdoctoral fellowship.

\subsection*{AI use disclosure}
During the preparation of this manuscript, the authors used OpenAI's ChatGPT-5.6 Sol to help identify patterns in the numerically computed singular values, particularly for the ladder networks, and to formalize the proofs presented in the appendix. The authors also used the tool to debug and improve the readability of the Python code and to assemble the package modules from existing functions. The authors reviewed and validated the AI-assisted outputs and take full responsibility for the manuscript.

\section*{Data Availability}
Interactive 3D network visualization in HTML format and additional figures are provided in the supplementary material. The Python code for generating the model networks studied here, constructing the ensnarlment operator, and performing the spectral and edge cut analyses is archived at 10.5281/zenodo.24591417.

\appendix
\section{Proofs}

\subsection{Reduction of the spectral problem to cycle space}\label{app:a1}

Consider an ensnarled system of two identical networks represented using corresponding edge and cycle coordinates, such that $M_1=M_2=M$. 
Let $M\in\mathbb{R}^{e\times c}$ have linearly independent columns and define $G=M^TM$. 
The ensnarlment operator is then
\[
    \L = MG^{-1}LG^{-1}M^T.
\]
Defining
\[
    A=MG^{-1},
    \qquad
    B=LG^{-1}M^T,
\]
gives $AB = \L$, while
\[
    BA = LG^{-1}M^TMG^{-1} = LG^{-1}.
\]
Sylvester's determinant identity gives
\begin{equation}
    \det(\lambda I_m-AB) = \lambda^{m-n} \det(\lambda I_n-BA).
\end{equation}
It follows that the nonzero eigenvalues satisfy
\begin{equation}
    \operatorname{spec}_{\ne 0}(\L) = \operatorname{spec}_{\ne 0}(LG^{-1}).
\end{equation}

In the case where the ensnarlment is such that $L^2 = I$, the nonzero eigenvalues of $\L$ can be further reduced as
\begin{equation} \label{eq11}
\operatorname{spec}_{\neq 0}(\L) = \frac{1}{\operatorname{spec}_{\neq 0}(LG)} .
\end{equation}

The singular-value problem admits a closely related reduction. Define
\[
    Q=MG^{-1/2},
    \qquad
    \widetilde{\L}=G^{-1/2}LG^{-1/2}.
\]
Since $Q^TQ=I$, the ensnarlment operator satisfies
\[
    \L = Q\widetilde{\L}Q^T,
\]
and hence
\[
    \L^T \L = Q\widetilde{\L}^T\widetilde{\L}Q^T.
\]
A further application of Sylvester's theorem gives
\[
    \operatorname{spec}_{\neq0}(\L^T\L) = \operatorname{spec}_{\neq0} \left( \widetilde{\L}^T\widetilde{\L} \right).
\]
Therefore,
\[
    \sigma_{\neq0}(\L) = \sigma_{\neq0} \left( G^{-1/2}LG^{-1/2} \right).
\]

%%%%%%%%%%%%%

\subsection{Rank lower bound for unlinking} \label{app:rank_lower_bound}

The rank lower bound established in Sec.~\ref{sec:edge_cutting} follows from the following auxiliary result. 

\begin{lemma}
Let $T:V \rightarrow W$ be a linear map of rank $r$, and let
$U\subseteq V$ have codimension $d$. Then
\[
    \operatorname{rank}(T|_U)\ge r-d\, ,
\]
where $T|_U : U \rightarrow W$ denotes the restriction of $T$ to $U$.
\label{lem:codimension}
\end{lemma}

\begin{proof}
Choose a complementary subspace $Z$ such that
\[
    V=U\oplus Z, \qquad \dim \left(Z\right)=d.
\]
Then
\[
    T(V)=T(U)+T(Z)\, .
\]
We further have that 
\[
    \operatorname{rank} (T) = \operatorname{dim} \left(T(V)\right),
    \qquad 
    \operatorname{rank} (T|_U) = \operatorname{dim}\left(T(U)\right),
\]
and therefore
% \[
%     \operatorname{rank}(T) \le \operatorname{rank}(T|_U) + \dim\left( T(Z)\right) \le \operatorname{rank}(T|_U) + d.
% \]
% \[
%     \operatorname{rank}(T|_U) \ge \operatorname{rank}(T)-d.
% \]
\begin{align}
    &\operatorname{rank}(T) \le \operatorname{rank}(T|_U) + \dim\left( T(Z)\right) \le \operatorname{rank}(T|_U) + d \nonumber\\
    &\Leftrightarrow \operatorname{rank}(T|_U) \ge \operatorname{rank}(T)-d \, .
\end{align}
\end{proof}

\begin{theorem}
Let $\Lambda$ be the ensnarlment operator associated with two ensnarled networks, then
\begin{equation}
    \operatorname{rank}(\Lambda) =\operatorname{rank}(L)\, .
\end{equation}
Let $r:=\operatorname{rank}(\Lambda) =\operatorname{rank}(L)$, then
% and let
% \[
% r:=\operatorname{rank}(\Lambda) =\operatorname{rank}(L).
% \]
any set of edge removals that completely unlinks the two networks contains at least $r$ edges.
\end{theorem}

% \cs{summarize proof, push proof to appendix. keep assumptions and final results}

\begin{proof}
Let $M_i\in\mathbb{R}^{E_i\times C_i}$ denote the mesh matrices whose columns form bases of the cycle spaces, and let
\[
    \Lambda=M_1G_1^{-1}LG_2^{-1}M_2^T,
    \qquad
    G_i=M_i^TM_i.
\]
%Since 
$M_iG_i^{-1}$ has full column rank by the definition of cycle bases, therefore, it follows that
\[
    \operatorname{rank}(\Lambda) = \operatorname{rank}(L) =:r \, .
\]

Suppose an edge $e$ is removed from the first network. Let
$m_e^T\in\mathbb{R}^{1\times C_1}$
denote the corresponding row of $M_1$. A cycle $x$ survives the cut if and only if it does not traverse the removed edge, i.e.,
\begin{equation*}
     \left(m_e^T\right)_x=0 \, ,
\end{equation*}
where $(\cdot)_x$ returns the $x$-th element of the vector.
Hence the surviving cycle space is
\[
    C_1 ^\prime =\ker(m_e^T),
\]
whose codimension satisfies
\[
    \operatorname{codim}(C_1 ^\prime) = \operatorname{rank}(m_e^T) \le 1.
\]

% We require the following lemma:

% \begin{lemma}
% Let $T:V\rightarrow W$ be a linear map of rank $r$, and let
% $U\subseteq V$ have codimension $d$. Then
% \[
%     \operatorname{rank}(T|_U)\ge r-d.
% \]
% \end{lemma}

% \begin{proof}
% Choose a complementary subspace $Z$ such that
% \[
%     V=U\oplus Z, \qquad \dim Z=d.
% \]
% Then
% \[
%     T(V)=T(U)+T(Z)
% \]
% We further have that 
% \[
%     \operatorname{rank} (T) = \operatorname{dim} T(V),
%     \qquad 
%     \operatorname{rank} (T|_U) = \operatorname{dim}T(U)
% \]
% and therefore
% \[
%     \operatorname{rank}(T) \le \operatorname{rank}(T|_U) + \dim T(Z) \le \operatorname{rank}(T|_U) + d.
% \]
% Rearranging gives
% \[
%     \operatorname{rank}(T|_U) \ge \operatorname{rank}(T)-d.
% \]
% \end{proof}

Since removing a single edge produces a restriction of codimension at most $1$, the rank of the residual linkage operator can decrease by at most $1$, by Lemma \ref{lem:codimension}.

Now let $X_1$ denote a set of $\chi_1$ removed edges from the first network, and define
\[
    A_{X_1} := (M_1)_{X_1}, \quad A_{X_1} \in \R^{\chi_1 \times C_1},
\]
where $(M_1)_{X_1}$ denotes the submatrix of $M_1$ consisting of rows indexed by $X_1$.

The surviving cycle space is
\[
    C_1 ^\prime = \ker(A_{X_1}),
\]
with
\[
    \operatorname{codim}(C_1 ^\prime) = \operatorname{rank}(A_{X_1}) \le \chi_1.
\]
Let $Q_1$ contain a basis for $C_1 ^\prime$. If no edges are removed from the second network, the residual linkage matrix is
\[
    L_{\mathrm{res}} = Q_1^TL \, .
\]
Applying Lemma \ref{lem:codimension} yields
\[
    \operatorname{rank}(L_{\mathrm{res}}) \ge r-\chi_1 \, .
\]

Complete unlinking requires
\[
    L_{\mathrm{res}}=0 \, ,
\]
so that
\[
    0 = \operatorname{rank}(L_{\mathrm{res}}) \ge r-\chi_1 \, ,
\]
which implies
\[
    \chi_1\ge r \, .
\]
% \end{proof}

% \begin{corollary}
% Let $X_1$ and $X_2$ denote sets of removed edges from the first and second networks, with
% $|X_1|=\chi_1$ and $|X_2|=\chi_2$. Then
% \[
%     \chi_1+\chi_2 \ge r \, ,
% \]
% where $r=\operatorname{rank}(\Lambda)$.
% \end{corollary}

% \begin{proof}
The similar idea applies to removing edges from both networks. Let $Q_1$ and $Q_2$ span the surviving cycle spaces after removing edges from the two networks. The residual linkage matrix is
\[
    L_{\mathrm{res}} = Q_1^TLQ_2 \, .
\]
Restricting the first cycle space lowers the rank by at most $\chi_1$, while restricting the second lowers it by at most $\chi_2$. Hence
\[
    \operatorname{rank}(L_{\mathrm{res}}) \ge r-\chi_1-\chi_2 \, .
\]
Since complete unlinking requires $L_{\mathrm{res}}=0$,
\[
    0 = \operatorname{rank}(L_{\mathrm{res}}) \ge r-\chi_1-\chi_2 \, ,
\]
which immediately gives
\[
    \chi_1+\chi_2\ge r \, .
\]
\end{proof}
\section{Analytical spectral results}

\subsection{Ladder spectrum derivation}\label{app:b1}

We consider ensnarled ladder graphs that are identical with period $p$, such that each 4-cycle of one graph is Hopf-linked to its complementary 4-cycle in the other graph (\figurename~\ref{fig:ladders_singvals}). 
In this case, the linkage matrix $L$ is diagonal; the Gram mesh matrices $G$ are Toeplitz tridiagonal of the form $\operatorname{tridiag}(a,b,c)$ where $a$ is the main diagonal and $b,c$ are the upper and lower diagonal elements respectively :
\[
L = \operatorname{diag}(1, -1, 1, -1, \ldots),\ G = \operatorname{tridiag}(4, -1, -1).
\]
By Eq.(\ref{eqn: sylvester_reduction}), the nonzero singular values of $\L$ are the inverse square roots of the eigenvalues of $(LG)^2$. 

Define $K = \operatorname{tridiag}(0,-1,-1)$. Then
\[
L^2 = I,\qquad G = 4I + K,\qquad LG = 4L + LK.
\]
We further have 
\[
LKL = -K,\qquad LK = -KL.
\]
Consequently,
\begin{align*}
(LG)^2 &= (4L + LK)^2  = 16L^2 + (LK)^2 + 4L(LK) + 4(LK)L
\\
&= 16I - K^2,
\\
\operatorname{spec}_{\neq 0}((LG)^2) 
&= \operatorname{spec}_{\neq 0}(16I - K^2) 
% = 16 - \operatorname{spec}_{\neq 0}(K^2) 
= 16 - (\operatorname{spec}_{\neq 0}(K))^2.
\end{align*}

The eigenvalues of the tridiagonal Toeplitz matrix $K$ follow from the closed form spectrum given in Eq.(\ref{eqn: toeplitz_tridiagonal_spectrum}). 
% $K$ is a Toeplitz tridiagonal matrix that has a closed-form solution for its spectrum \cite{smith1985numerical}: 
% \[
% spec(tridiag(a,b,c)) = a - 2\sqrt{bc}\cos\left({\frac{k\pi}{n+1}}\right) ,\ k = 1(1)n
% \]
\begin{align*}
\operatorname{spec}((LG)^2) &= 16 - 4 \cos^2 \left( {\frac{k\pi}{p+1}} \right) ,\qquad k = 1,2, \ldots, p.
\end{align*}

These eigenvalues are invariant under the index transformation $k \rightarrow n+1-k$, since 
\[
    \cos^2 \left( {\frac{k\pi}{p+1}} \right) = \cos^2 \left( {\frac{(p+1-k)\pi}{p+1}} \right).
\]

Thus, the modes indexed by $k$ and $p+1-k$ have the same eigenvalue. 
For even $p$, these indices form distinct pairs, and hence every spectral value is two-fold degenerate. 
For odd $p$, all indices form distinct pairs except the fixed point  
\[
    k = \frac{p+1}{2},
\]
which produces a single nondegenerate eigenvalue.

Using $4\cos^2 x = 2 + 2\cos(2x)$, the distinct eigenvalues of $(LG)^2$ can be written as
\begin{align*}
\operatorname{spec}(LG)^2
&= 14 - 2 \cos \left( {\frac{2k\pi}{p+1}} \right)
\\
&= 14 + 2 \cos \left( \frac{\pi(p+1-2k)}{p+1} \right),
\qquad k = 1,2, \ldots, \lceil p/2\rceil .
\end{align*}

For even period $p=2m$,
\[
p+1-2k = 2j - 1,
\]
where $j = m-k+1$.

For odd period $p = 2m-1$,
\[
p+1-2k = 2(j-1).
\]

From Eq.(\ref{eqn: sylvester_reduction}), the singular values of $\L$ are the inverse square roots of the corresponding eigenvalues of $(LG)^2$. 
The final closed form singular value spectrum for $\L$ is therefore 
\begin{gather}
    \s(L) = \frac{1}{\sqrt{14 + 2 \cos \theta}} \nonumber,
    \\
    \text{where,}
    \\
    \theta = \begin{cases}
        \displaystyle{\frac{(2j-1)\pi}{p+1}} & \text{even } p,
        \\[5pt]
        \displaystyle{\frac{2(j-1)\pi}{p+1}} & \text{odd } p,
    \end{cases} \quad
    j = 1,2, \ldots,  \lceil p/2 \rceil.
    \label{eqn: app_ladder_spectra}
\end{gather}

% The degeneracy structure is inherited from the $k \rightarrow n+1-k$ symmetry described above. 
% For even $n$, all singular values occur in degenerate pairs. 
% For odd $n = 2m-1$, the nondegenerate mode corresponds to the fixed point $k = m = (n+1)/2$. 
% At this index, 
% \[
%     \cos \left( \frac{k \pi}{n+1} \right) = \cos \left( \frac{\pi}{2} \right) = 0 
% \]
% and therefore the corresponding eigenvalue of $(LG)^2$ is
% \[
%     \l_k ( (LG)^2 ) = 16
% \]
% The corresponding singular value of $\L$ is consequently 
% \[
%     \s = \frac{1}{\sqrt{16}} = \frac{1}{4}.
% \]
% Since 
% \[
%     16 - 4 \cos^2 \left( \frac{k \pi}{n+1} \right) \leq 16, 
% \]
% this mode corresponds to the smallest singular value. 

\subsection{Principal singular value saturation for grids} \label{app:b2}

As shown in Sec.~\ref{sec:grids}, the principal singular value increases with the grid period and approaches 
\[
    \lim_{p \rightarrow \infty} \s_1 (p) = \frac{1}{2}.
\]
To understand this limit, we exploit the regular matrix structure of the grids. 
Since the two grids have identical mesh Gram matrices $G$, as shown in Appendix~\ref{app:a1}, the nonzero singular values of the ensnarlment operator are those of the reduced matrix 
\[
    \widetilde{\L} = G^{-1/2} L G^{-1/2}.
\]
For a $p \times p$ grid, 
\[
    G = 4I - A_x - A_y,
\]
where $A_x$ and $A_y$ couple horizontally and vertically adjacent 4-cycle plaquettes. 
The linkage matrix can be written as 
\[
    L = D (I - S_{xy}),
\]
where $D$ contains the alternating linkage signs and $S_{xy}$ shifts a plaquette by one position along both grid directions. 

In the infinite grid limit, a plane wave mode with wave vector $k_x,k_y$ gives
\[
    g(k_x,k_y) = 4 - 2 \cos k_x - 2\cos k_y
\]
for the gram matrix. The alternating signs in $D$ shift the wave vector by $(\pi, \pi)$, while $I-S_{xy}$ contributes the factor $1 - e^{-i(k_x + k_y)}$. 
The corresponding singular value function is therefore 
\[
    \s(k_x,k_y) = \frac{\left|1-e^{-i(k_x+k_y)}\right|}{\sqrt{g(k_x,k_y)\, g(k_x+\pi,k_y+\pi)}}.
\]
Introducing
\[
    u = \frac{k_x + k_y}{2}, \qquad v = \frac{k_x - k_y}{2},
\]
the singular value function becomes 
\[
    \s(u,v) = \frac{|\sin u|}{2 \sqrt{1 - \cos^2 u \, \cos^2 v}}.
\]
Since 
\[
    1 - \cos^2 u \, \cos^2 v = \sin^2 u + \cos^2 u \, \sin^2 v \geq \sin^2 u,
\]
we obtain 
\[
    \s(u,v) \leq \frac{1}{2}.
\]
The bound is attained for $v=0$, corresponding to $k_x = k_y$, for $\sin u \neq 0$, and hence the principal singular value converges to $1/2$ as the finite grid approaches the translation invariant limit.

\section{Periodic boundary effects} \label{app:c}

Here we consider a periodic extension of the ladder networks, where we close the two ends of the ladders. 
We choose their embeddings such that one network forms a planar prism graph, while the other forms a 3d prism graph. 
Each 4-cycle in a network is Hopf-linked to two 4-cycles in the other network. 
Although the two networks have different spatial embeddings, they share the same topology. 
Relative to open ladders, these closed prisms have one additional independent basis cycle formed by the rail edges.

The resulting singular value spectra exhibit a periodic structure. 
For a prism of period $p$, the $p-1$ singular values are 
\[
    \s(p) = \frac{ \sin \left( \pi k/p \right) }{ 2-\cos \left( 2\pi k/p \right) },
    \qquad
    k=1,\ldots,p-1.
\]

The symmetry $\s_k = \s_{p-k}$ produces twofold degenerate singular values. 
For odd $p$, the values therefore occur in degenerate pairs, while for even $p$ there is one nondegenerate value
\[
    \s_{p/2} = \frac{1}{3},
\]
occurring at the mode $k=p/2$.

When $p$ is a multiple of 6, the modes $k = p/6$, $p/2$, and $5p/6$ coincide at $\s = 1/3$, producing a threefold degeneracy. 

The associated singular subspaces show a consistent separation between rung and rail dominated modes. 
From the principal through the intermediate  subspaces, the rung edges have larger relative contribution than rail edges, although their numerical weights vary between the modes. 
This pattern reverses in the minor subspace, where the rail edges have higher scores. 

\begin{figure}[h]
    \centering
    \includegraphics[width=0.8\linewidth]{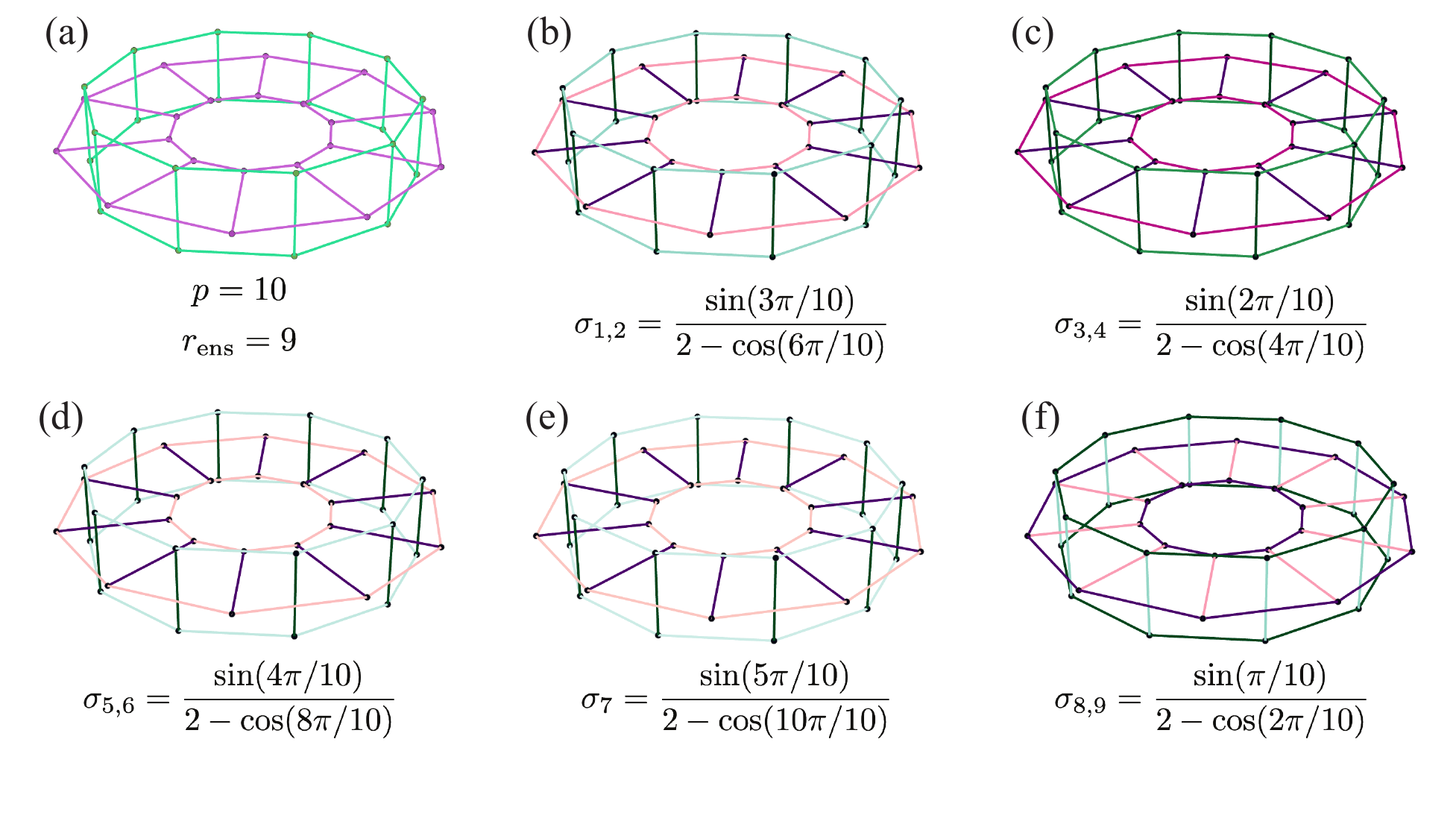}
    \caption{
    Periodic boundary effects in ensnarled ladders.
    (a) A pair of ensnarled ladders with their longitudinal ends identified to form closed rings, shown for period $p=10$ and ensnarlment rank $r_{\mathrm{ens}}=9$.
    (b-f) Spatial distributions of the left and right ensnarlment centralities for the singular subspaces, ordered from the principal to the minor singular value.
    The two colors distinguish the networks, and the visual weight of each edge represents its ensnarlment centrality.
    (b) Principal two-dimensional singular subspace associated with $\sigma_{1,2}$.
    (c-e) Intermediate singular subspaces associated with $\sigma_{3,4}$, $\sigma_{5,6}$, and the nondegenerate $\sigma_7$, respectively.
    (f) Minor two-dimensional singular subspace associated with $\sigma_{8,9}$.
    Closing the longitudinal boundaries removes the central-to-peripheral organization of the open ladders: the higher singular subspaces preferentially weight the rungs shared by adjacent cycles, whereas the minor subspace weights the rail edges.
    The two color families distinguish
    the two networks. 
    Edge darkness encodes the magnitude of the corresponding component of the unit-normalized singular vector: darker edges have larger component magnitudes and participate more strongly in that singular mode. 
    Color intensity should therefore be interpreted relative to other edges within the same displayed vector.
    }
    \label{fig:ringladders}
\end{figure}

This low resolution of spatial hierarchy stratified by the singular subspace vectors hints at the role of boundary conditions of the system.
Once the ends are identified, no position along the ladder is geometrically distinguished, and we no longer observe a localized set of edges that can be regarded as truly ``central''.
Instead, equivalent edges recur periodically around the ring. 
This suggests that the spatial localization of edges in the open ladders, and other bounded network archetypes shown through Sec.~\ref{sec:spectral_properties}, is a finite boundary effect.

\newpage
\bibliographystyle{unsrt}
\bibliography{bib}

% \cs{
% Notation:
% \begin{itemize}[noitemsep]
%     \item scalars: lowercase
%     \item vectors: lowercase bold 
%     \item operators/matrices: uppercase
%     \item sets: mathcal uppercase
% \end{itemize}
% }

\end{document}